\documentclass[oneside,12pt,reqno]{amsart}
\usepackage{amssymb, amscd}
\usepackage[all]{xy}
\usepackage{graphics}
\usepackage{textcomp}
 \newtheorem{theorem}{Theorem}

 \newtheorem{proposition}[theorem]{Proposition}
 \newtheorem{definition}[theorem]{Definition}
 \theoremstyle{definition}
 
 \theoremstyle{remark}
 \newtheorem{remark}[theorem]{Remark}

\def\noi{\noindent}

\def\al{\alpha}
\def\be{\beta}
\def\ga{\gamma}

\def\BZ{{\mathbb{Z}}}

\def\ss{\mbox{\boldmath ${s}$}}
\def\ssp{\mbox{\boldmath ${s'}$}}
\def\sspp{\mbox{\boldmath ${s''}$}}
\def\nn{\mbox{\boldmath ${n}$}}
\def\nnp{\mbox{\boldmath ${n'}$}}
\def\nnpp{\mbox{\boldmath ${n''}$}}

\def\ff{\mbox{\boldmath ${f}$}}

\newlength{\vscaling} \newlength{\hscaling}
\usepackage{latexsym}
\usepackage{amssymb}

\def\U{{\mathcal {U}}}

\def\GG{{\mathcal G}}

\def\NN{{\mathcal N}}
\def\LL{{\mathcal L}}
\def\EE{{\mathcal E}}
\def\TT{{\mathcal T}}
\def\Ii{{\mathcal I}}

\def\P2{{P^{[2]}}}
\def\vphi{{\varphi}}
\def\vvphi{\mbox{\boldmath $\vphi$}}
\def\vvpsi{\mbox{\boldmath $\psi$}}
\def\vvxi{\mbox{\boldmath $\xi$}}
\def\vvph{{\mbox{\boldmath $\scriptstyle \vphi$}}}
\def\vvps{{\mbox{\boldmath $\scriptstyle \psi$}}}

\def\ssigma{\mbox{\boldmath $\sigma$}}
\def\sssigma{{\mbox{\boldmath $\scriptstyle \sigma$}}}
\def\ttt{{\mbox{\boldmath $\scriptstyle t$}}}
\def\tt{{\mbox{\boldmath $t$}}}
\def\ll{{\mbox{\boldmath $\ell$}}}
\def\11{{\mbox{\boldmath $1$}}}
\def\YO{Y_{\,\,}\!|_{O}}
\def\eq{\begin{equation}}
\def\en{\end{equation}}
\def\sk{\vskip .4cm}

\def\eq#1\en{\begin{equation}#1\end{equation}}
\def\eqa#1\ena{\begin{eqnarray}#1\end{eqnarray}}
\def\zz{{\mbox{\boldmath $z$}}}
\def\ii{{\mbox{\boldmath $i$}}}

\def\V{\mbox{V}}
\def\H{\mbox{H}}

\newcommand{\+}{{\mbox{\boldmath $+_{}$}}}
\newcommand{\aalpha}{{\mbox{\boldmath ${\al}$}}}
\newcommand{\ax}{{\mbox{\boldmath $a$}}}
\newcommand{\bb}{{\mbox{\boldmath $b$}}}
\newcommand{\cc}{{\mbox{\boldmath $c$}}}
\newcommand{\ccs}{{\mbox{\boldmath $\scriptstyle c$}}}
\newcommand{\hh}{{\mbox{\boldmath $h$}}}
\newcommand{\axp}{{\mbox{\boldmath $a'$}}}

\newcommand{\axs}{{\mbox{\boldmath $\scriptstyle a$}}}
\newcommand{\kx}{{\mbox{\boldmath $k$}}}

\newcommand{\kxs}{{\mbox{\boldmath $\scriptstyle k$}}}
\newcommand{\omm}{{\mbox{\boldmath $\omega$}}}
\newcommand{\varthees}{{\mbox{\boldmath $\scriptstyle \vartheta$}}}
\newcommand{\Dd}{{\mbox{\boldmath $D$}}}
\newcommand{\Hh}{{\mbox{\boldmath $H$}}}
\newcommand{\chibold}{{\mbox{\boldmath $\chi$}}}
\newcommand{\tdeltt}{{\mbox{\boldmath $\tilde\delta$}}}
\newcommand{\thetaa}{{\mbox{\boldmath $\theta$}}}
\newcommand{\thetaas}{{\mbox{\boldmath $\scriptstyle \theta$}}}

\newcommand{\circs}{{{\scriptstyle{^{_{_{\scriptstyle{\circ}}}}}}}}
\newcommand{\varthee}{{\mbox{\boldmath $\vartheta$}}}

\newcommand{\ZZ}{{\mathcal Z}}
\newcommand{\QQ}{{\mathcal Q}}

\newcommand{\II}{\mathcal{I}}

\newcommand{\mlie}{\rm{Lie}}

\begin{document}
\begin{titlepage}
\vspace{1.cm}
\begin{flushright}
MPP-2003-139\\
LMU-TPW 07/03\\
\end{flushright}
\vspace{1 cm}

\begin{centering}
\vspace{.43in} 
{\Large {\bf Nonabelian Bundle Gerbes, their Differential Geometry and
  Gauge Theory}}\\ \vspace{1.5cm}

{\bf Paolo Aschieri},$^{1,2,3,a}$ {\bf Luigi Cantini}$^{4,5,b}$ and {\bf
Branislav Jur\v co}$^{2,3,c}$\\

\vspace{.25in}$^{1}$Dipartimento di Scienze e Tecnologie Avanzate\\ 
Universit\'a del Piemonte Orientale, and INFN\\ 
Corso Borsalino 54, I-15100,  Alessandria, Italy\\
 \vspace{.15cm}
$^{2}$Max-Planck-Institut f\"{u}r Physik\\ F\"{o}hringer Ring 6, D-80805
M\"{u}nchen\\ \vspace{.15cm}
$^{3}$Sektion Physik, Universit\"{a}t M\"{u}nchen\\
Theresienstr. 37, D-80333 M\"{u}nchen\\ \vspace{.15cm}
$^{4}$Scuola Normale Superiore\\ 
Piazza dei Cavalieri 7, 56126 Pisa \& INFN sezione di Pisa\\ \vspace{.15cm}
$^{5}$ 
Department of Physics\\
Queen Mary, University of London\\
       Mile End Road,
       London E1 4NS\vspace{.15cm}

\vspace{0.8in}

{\bf Abstract}\\
\end{centering}
\vspace{.1in}
 \noindent 
Bundle gerbes are a higher version of line bundles, we present
nonabelian bundle gerbes as a higher version of principal
bundles. Connection, curving, curvature and gauge transformations are
studied both in a global coordinate independent formalism and in 
local coordinates. These are the gauge fields needed for the
construction of Yang-Mills theories with 2-form gauge potential.

\vspace{3.cm}

\begin{flushleft}
$^{a}$e-mail address: aschieri@theorie.physik.uni-muenchen.de\\
$^{b}$e-mail address: l.cantini@qmul.ac.uk\\
$^{c}$e-mail address: jurco@theorie.physik.uni-muenchen.de 
\end{flushleft}
\end{titlepage}

\setcounter{page}{1}
\section{\bf Introduction}
Fibre bundles, besides being a central subject in geometry and
topology, provide the mathematical framework for describing
global aspects of Yang-Mills theories. Higher abelian gauge theories,
i.e. gauge theories with abelian 2-form gauge potential appear
naturally in string theory and field theory, and here too we have 
a corresponding mathematical structure,
that of abelian gerbe (in algebraic geometry) and of abelian bundle
gerbe (in differential geometry). Thus abelian bundle gerbes are a 
higher version of line bundles. 
Complex line bundles are geometric realizations of the integral $2$nd cohomology
classes $H^2(M,\BZ)$ on a manifold, i.e. the first Chern classes (whose 
de Rham representative is the field strength). 
Similarly, abelian (bundle) gerbes are the next level in realizing
integral cohomology classes on a manifold, they are geometric
realizations of the $3$rd cohomology classes
$H^3(M,\BZ)$. Thus the curvature 3-form of a 2-form gauge potential is the
de Rham representative of a class in $H^3(M,\BZ)$. This class is
called the Dixmier-Douady class \cite{Giraud},\cite{Brylinski:ab}; 
it topologically characterizes
the abelian bundle gerbe in the same way that the first Chern class
characterizes complex line bundles.

One way of thinking about abelian
gerbes is in terms of their local transition functions
\cite{Hitchin:1999fh},\cite{Chat}.
Local ``transition functions" of an abelian gerbe are complex line
bundles on double overlaps of open sets satisfying cocycle conditions  
for tensor products over quadruple overlaps of open sets.  
The nice notion of abelian bundle gerbe 
\cite{Murray} is related to this picture. 
Abelian gerbes and bundle gerbes can be equipped with
additional structures, that of connection 1-form and of curving (the
2-form gauge potential), and that of (bundle) gerbe modules (with or without
connection and curving ). Their holonomy can be introduced and studied
\cite{Hitchin:1999fh},\cite{Chat},\cite{Bouwknegt:2001vu},\cite{Carey:2002xp},\cite{Mackaay},\cite{MP1}. 
The equivalence class of an abelian gerbe with connection and curving
is the Deligne class on the base manifold. The top part of the Deligne
class is the class of the curvature, the Diximier-Douady class.

Abelian gerbes arise in a natural way in 
quantum field theory \cite{CMM1},\cite{CMM2},\cite{Carey:2001gi},
where their appearance
is due to the fact that one has to deal with abelian extensions of the group
of gauge transformations; this is related to chiral anomalies. 
Gerbes and gerbe modules
appear also very naturally in TQFT \cite{Picken}, in the WZW model 
\cite{Gawedzki:2002se} and in the description of D-brane anomalies in 
nontrivial background 3-form
$H$-field (identified with the Diximier-Douday class)
\cite{Freed:1999vc},\cite{Kapustin:1999di},\cite{Bouwknegt:2000qt}. Coinciding (possibly infinitely many) D-branes are submanifolds ``supporting''
bundle gerbe modules \cite{Bouwknegt:2001vu} and can be classified by
their (twisted) $K$-theory. The relation to the boundary conformal field
theory description of D-branes is due to the identification
of equivariant twisted $K$-theory with the Verlinde algebra
\cite{Mickelsson},\cite{FHT}.
For the role of $K$-theory in D-brane physics
see e.g. \cite{Witten:1998cd},\cite{Witten:2000cn},\cite{Szabo}.

\sk
In this paper we study the nonabelian generalization of abelian bundle gerbes and
their differential geometry, in other words we study higher
Yang-Mills fields.  Nonabelian gerbes arose in the
context of nonabelian cohomology
\cite{Dedecker},\cite{Giraud} (see \cite{Moerdijk} for a concise
introduction), see also (\cite{266}). Their differential geometry --from the
algebraic geometry point of view-- is discussed thoroughly in the recent
work of Breen and Messing \cite{Breen-Messing} (and their combinatorics
in \cite{Attal}). Our study on the other hand is from the differential
geometry viewpoint. We show that nonabelian bundle gerbes connections and curvings
are very natural concepts also in classical differential geometry.  
We believe that it is primarily in this context that these
structures can appear and can be recognized in physics. 
It is for example in this context that one would like to
have a formulation of Yang-Mills theory with higher forms.
These theories should be relevant in order to describe 
coinciding NS5-branes with D2-branes ending on them. 
They should be also relevant in the study of M5-brane anomaly.
We refer to \cite{Kalkkinen:1999pm}, \cite{Baez}, \cite{Hofman} for 
some attempts in constructing higher gauge fields.

\sk

Abelian bundle gerbes are constructed using  
line bundles and their products. 
One can also  study $U(1)$ bundle gerbes, here line bundles are replaced by
their corresponding principal $U(1)$ bundles. In the study of  
nonabelian bundle gerbes it is more convenient to work with nonabelian
principal bundles then with vector bundles. 
Actually principal bundles with additional structures are needed. 
We call these objects (principal) bibundles and $D$-$H$ bundles ($D$ and $H$ being
Lie groups). 
Bibundles are fibre bundles (with fiber $H$) which are at the same time
left and right principal bundles (in a compatible way). They are the
basic objects for constructing (principal) nonabelian bundle gerbes.  
The first part of this paper is therefore devoted to their description. In 
Section 2 we introduce bibundles, $D$-$H$ bundles (i.e. principal $D$
bundles with extra $H$ structure) and study their products.
In Section 3 we study the differential geometry of bibundles, in particular we
define  connections, covariant exterior derivatives and
curvatures. These structures are generalizations of the corresponding structures
on usual principal bundles.  We thus describe them using a language
very close to  that of the classical reference books 
\cite{Koba-Nomizu} or \cite{Husemoller}. 
In particular a connection on a bibundle needs to satisfy a relaxed 
equivariance property, this is the price to be paid in order to
incorporate nontrivially the additional bibundle structure. 
We are thus lead to introduce the notion of a $2$-connection $(\ax, A)$ on a bibundle. 
Products of bibundles with connections give a bibundle with connection
only if the initial connections were compatible, we call this
compatibility the summability conditions for $2$-connections; a
similar summability condition is established also for 
horizontal forms (e.g. $2$-curvatures). 

In Section $4$, using the product between bibundles 
we finally introduce (principal) bundle gerbes. Here too we first describe
their  structure (including stable equivalence) and then only later in
Section 7  we describe their differential geometry. 
We start with the proper generalization of abelian bundle gerbes in
the sense of Murray \cite{Murray}, we then describe the relation to the
Hitchin type presentation \cite{Hitchin:1999fh},\cite{Chat}, where
similarly to the abelian case, nonabelian gerbes are described
in terms of their "local transition functions" which are bibundles
on double overlaps of open sets. The properties of the products of
these bibundles over triple and quadruple overlaps define the 
gerbe and its nonabelian \v{C}ech 2-cocycle.

Section $5$ is devoted to the example of the lifting bundle gerbe
associated with the group extension $H\rightarrow E\rightarrow G$.  
In this case the bundle gerbe with structure group $H$
appears as an obstruction to lift to $E$ a $G$-principal bundle $P$. 

Again by generalizing the abelian case, bundle gerbe modules are
introduced in Section
$6$. Since we consider principal bibundles we obtain modules that 
are $D$-$H$ bundles (compatible with the bundle gerbe structure). With each bundle
gerbe there is canonically associated an
$Aut(H)$-$H$ bundle.
In the lifting bundle gerbe example a module is given by the 
trivial $E$-$H$ bundle.

In Section $7$ we introduce the notion of bundle gerbe connection and prove
that on a bundle gerbe a connection always exists. 
Bundle gerbe connections are then equivalently  described as collections of local 
$2$-connections on local bibundles (the ``local transition
functions of the bundle gerbe'') satisfying a nonabelian cocycle condition
on triple  overlaps of open sets. Given a bundle gerbe connection we
immediately have a connection on the canonical bundle gerbe module $can$. We
describe also the case of a bundle gerbe connection associated with an arbitrary
bundle gerbe module. In particular we describe the bundle gerbe connection in
the case of a lifting bundle gerbe.

Finally in Section $8$ we introduce the nonabelian curving $\bb$ (the
2-form gauge potential) and the 
corresponding nonabelian curvature $3$-form  $\hh$.
These forms are the nonabelian generalizations of the string theory $B$ and $H$ 
fields. 
\sk
\sk

\sk
\section{\bf Principal Bibundles and their Products}
\label{sec2}
Bibundles (bitorsors) where first studied by Grothendieck \cite{Gro} and Giraud
\cite{Giraud}, their cohomology was studied in \cite{Breen}.
We here study  these structures using the language of
differential geometry.

Given two $U(1)$ principal bundles $E$, $\tilde E$, on the same base
space $M$, one can consider the fiber product bundle  $E\tilde E$, 
defined as the
$U(1)$ principal bundle on $M$ whose fibers are the product of the $E$ 
and $\tilde E$, fibers. If we introduce a local description of $E$ and  
$\tilde E$, with transition functions $h^{ij}$ and $\tilde h^{ij}$
(relative to the covering $\{U^i\}$ of $M$), 
then $E\tilde E$ has transition functions $h^{ij}\tilde h^{ij}$.

In general, in order to multiply principal nonabelian bundles 
one needs extra structure. Let $E$ and $\tilde E$ be $H$-principal
bundles, we use the convention that $H$ is acting on the bundles from the left. 
Then in order to define the $H$ principal left bundle $E\tilde E$ we need also a
right action of $H$ on $E$. We thus arrive at the following
\begin{definition}

An $H$ principal bibundle $E$ on the base space $M$
is a bundle on $M$ that is both a  {\sl left} $H$
principal bundle 
and a {\sl right}  $H$ principal bundle and where left and right $H$ actions
commute 
\eq
\forall\;h,k\in H ~,~\forall e\in E,~~~~~~
(k \,e)\triangleleft h=k_{}( e\triangleleft h)~;
\en
we denote with $p\;:E\rightarrow M$ the projection to the base space.
\end{definition}

Before introducing the product between principal bibundles we briefly
study their structure. 
A morphism $W$ between two principal bibundles $E$ and $\tilde E$ 
is a morphism between the bundles $E$ and $\tilde E$ compatible with 
both the left and the right action of $H$:
\eq
W (k_{\:}e\triangleleft h)=k_{\:}W(e)_{}\tilde\triangleleft_{\,}h~,
\label{isoal}
\en
here $\tilde\triangleleft$ is the right action of $H$ on $\tilde E$.
As for morphisms between principal bundles on the same base space $M$, 
we have that morphisms between principal bibundles on $M$ are isomorphisms.

\subsection*{Trivial bibundles} 
\noindent \\
Since we consider only principal bibundles we will frequently write
bibundle for principal bibundle.
The product bundle $M\times H$ where left and right actions are
the trivial ones on $H$ [i.e. $k_{\,}(x,h)\triangleleft h'=(x,khh')$] is a
bibundle. We say that a  bibundle $T$ is trivial if
$T$ is isomorphic to $M\times H$.
\begin{proposition}
We have that $T$ is trivial as a bibundle iff it has a global
central section, i.e. a global section 
$\ssigma$ that intertwines the left and the right action of $H$ on $T$: 
\eq
\forall h\in H\,,\,\forall x\in M,     ~~      ~~
h_{\,}\ssigma(x)=\ssigma(x)\triangleleft h\label{left&rightiso}
\en
\end{proposition}
\begin{proof}Let $\ssigma$ be a global section of $T$, define 
$W_{\!\sssigma}\,:\,M\times H\rightarrow T$ as
$W_{\!\sssigma}(x,h)=h_{\,}
\ssigma(x)$,
then $T$ and $M\times H$ are isomorphic as left principal bundles. The
isomorphism $W_{\!\sssigma}$ is also a right principal bundles isomorphism iff 
(\ref{left&rightiso}) holds.
\end{proof}
Note also that the section $\ssigma$ 
is unique if $H$ has trivial centre. An example of nontrivial bibundle
is given by the trivial left bundle $M\times H$ equipped with the
nontrivial right action $(x,h)\triangleleft h'=(x,h\chi(h'))$ where
$\chi$ is an outer automorphism of $H$.  
We thus see that bibundles are in general {\sl not} locally trivial. 
Short exact sequences of groups provide examples of
bibundles that are in general nontrivial as left bundles
[cf. (\ref{ext}), (\ref{109})].

\subsection*{The $\vvphi$ map}
\noindent \\
We now further characterize the relation between left and right actions.
Given a bibundle $E$, the map $\vvphi\,:\,E\times H\rightarrow H$ defined by
\eq
\forall e\in E\,,~\forall h\in H\,,~~~~\vvphi_e(h)_{\,}e=e\triangleleft h
\label{rightaction}
\en
is well defined because the left action is free, and transitive on the fibers.
For fixed $e\in E$ it is also one-to-one since the right action is
transitive and left and right actions are free.
Using the compatibility between left and right actions it is not
difficult to show that $\vvphi$ is equivariant w.r.t. the left action
and that for fixed $e\in E$ it is an automorphism of $H$:
\eqa
&&\vvphi_{h_{}e}(h')=h_{}\vvphi_e(h')_{}h^{-1}\label{equivariance}~,\\
&& \vvphi_e(hh')=\vvphi_e(h)\vvphi_e(h')\label{automorphism}~,
\ena
we also have 
\eqa
&&\vvphi_{e\triangleleft h}(h')=\vvphi_e(hh'h^{-1})~.\label{extra}
\ena
Vice versa given a left bundle $E$ with an equivariant map $\vvphi \,:\,E\times
H\rightarrow H$ that restricts to an $H$ automorphism $\vvphi_e$,
we have that 
$E$ is a bibundle with right action defined by (\ref{rightaction}).

Using the $\vvphi$ map we have that a global section $\ssigma$ is a
global central section (i.e. that a trivial left principal
bundle is  trivial as bibundle) iff
[cf. (\ref{left&rightiso})],
$\forall x\in M$ and $\forall h\in H\,,$
\eq
\vvphi_{\sssigma(x)}(h)=h ~.
\label{invariance}
\en
In particular, since $e\in E$ can be always written as $e=h'\ssigma$, 
we see that $\vvphi_e$ is always an {\sl inner} automorphism, 
\eq
\vvphi_e(h)=\vvphi_{h'\sssigma}(h)=Ad_{h'}(h)~.
\en
Vice versa, we have that 
\begin{proposition} 
If $H$ has trivial centre then an $H$ 
bibundle $E$ is trivial iff
$\vvphi_e$ is an inner automorphism for all $e\in E$. 
\end{proposition}
\begin{proof}
Consider the local sections 
$\tt^i\,:\,U^i\rightarrow E$, since $H$ has trivial centre the map
$k(\tt)\,:\,U^i\rightarrow H$ is uniquely defined by 
$\vvphi_{\ttt^i}(h')=\,Ad_{k({\ttt^i})}h'$. 
{}From (\ref{equivariance}),
$\vvphi_{h\ttt^i}(h')=
Ad_h\,Ad_{k({\ttt^i})}h'$ and therefore the sections
${k(\tt^i)}^{-1}\tt^i\,$ are 
central because they  satisfy $\vvphi_{{k(\ttt^i)}^{-1}\ttt^i}(h')=h'\,$.
In the intersections $U^{ij}=U^i\cap U^j$ 
we have  $\tt^i=h^{ij} \tt^j$ and therefore
${k(\tt^i)}^{-1}\tt^i={k(\tt^j)}^{-1}\tt^j$.
We can thus construct a global central section.
\end{proof}

Any principal bundle with $H$ abelian is a principal bibundle in a trivial
way, the map $\vvphi$ is given simply by $\vvphi_e(h)=h$. 

Now let us recall that a global section $\ssigma : M\rightarrow E$ on a
principal $H$-bundle 
$E\rightarrow M$ can be identified with an $H$-equivariant map 
$\overline{\sigma}: E \rightarrow H$. With
our (left) conventions, $\forall E\in E$, 
$$
e ={\overline\sigma} (e)\ssigma (x)~.
$$
Notice, by the way, that if  $E$ is a
trivial bibundle with a global section $\ssigma$,
then $\overline{\sigma}$ is bi-equivariant, i.e.:
${\overline\sigma} (heh')=h{\overline\sigma} (e)h'$ iff $\sigma$ is central.
We apply this description of a global section of a left principal bundle  to
the following situation. Consider an $H$-bibundle $E$. Let us form
$Aut(H)\times_H E$ with the help of the canonical homomorphism $Ad : H
\rightarrow Aut(H)$. Then it is 
straightforward to check that $\overline{\sigma} : [\eta, e] \mapsto  \eta \circ \vvphi_e$
with $\eta \in Aut(H)$ is a global section of the left $Aut(H)$-bundle
$Aut(H) \times_H E$. So $Aut(H) \times_H E$ is trivial as a left 
$Aut(H)$-bundle. 
On the other hand if $E$ is a left principal $H$-bundle such that
$Aut(H) \times_H E$ is a trivial left $Aut(H)$-bundle then it has a global
section $\overline{\sigma} : Aut(H) \times_H E
\rightarrow Aut(H)$ and the structure of an $H$-bibundle
on $E$ is given by $\vvphi_e \equiv \overline{\sigma}([id,e])$. 
We can thus characterize $H$-bibundles without mentioning their 
right $H$ structure,
\begin{proposition}\label{4}
A left $H$-bundle $E$ is an $H$-bibundle if and only if  the (left) $Aut(H)$-bundle 
$Aut(H) \times_H E$ is trivial.
\end{proposition}

Any trivial left $H$-bundle $T$ can be given a trivial
$H$-bibundle structure. We consider a trivialization of $T$ i.e.
an isomorphism $T \rightarrow M\times H$ and pull back the trivial right
$H$-action on $M\times H$ to $T$. This just
means that the global section of the left $H$-bundle $T$ associated with
the trivialization $T \rightarrow M\times H$, is by definition
promoted to a global central section. 

\subsection*{Product of bibundles}
\noindent \\
In order to define the product bundle $E\tilde E$ we first consider the
fiber product (Withney sum) bundle 
\eq
E\oplus\tilde E\equiv\{(e,\tilde e)~|~p(e)=\tilde p(\tilde e) \}
\en
with projection $\rho\;:E\oplus\tilde E\rightarrow M$ given by
$\rho(e,\tilde e)=p(e)=\tilde p(\tilde e)$.
We now can define the product bundle $E\tilde E$ with base space $M$
via the equivalence relation
\eq
\label{12}
\forall h\in H~~~~
(e,h\tilde e)\sim (e\triangleleft h\,,{\tilde e})
\en
we write $[e,\tilde e]$ for the equivalence class and 
\eq
\label{13}
E\tilde E \equiv \,E\oplus_{H}\tilde E\equiv
\{_{\,}[e,\tilde e]_{\,}\}
\en
the projection 
$p\tilde p\;:\;E\tilde E\rightarrow M$ is given by $p\tilde p
[e,\tilde e]=p(e)=\tilde p (\tilde e)$.
One can show that $E\tilde E$ is an $H$ principal bundle; the
action of $H$ on  $E\tilde E$ is inherited from that on $E$:
$h [e,\tilde e]=[h_{}e,\tilde e]$.
Concerning the product of sections we have that if
$\ss ~:~U\rightarrow E$ is a section of $E$ 
(with $U\subseteq M$), and $\tilde{\ss}~:~U\rightarrow \tilde E$ is 
a section of $\tilde E$, then 
\eq
\ss\mbox{\boldmath $\tilde s$}\equiv[\,\ss,\mbox{\boldmath $\tilde s$}\,]~: ~U\rightarrow E\tilde E \label{sproduct}
\en
is the corresponding section of $E\tilde E$.

\sk
When also  $\tilde E$ is an $H$ principal bibundle, with right action
$\tilde{\triangleleft}$, then $E\tilde E$ is again an $H$ principal
bibundle with right action $\triangleleft_{\!}\tilde\triangleleft$ given by
\eq
[e,\tilde e]\triangleleft_{_{\!\!}}\!\tilde\triangleleft{\,} h\,=\,
[e,\tilde e_{\,}\tilde\triangleleft_{\,} h]
\en
It is easy to prove that the product between $H$ 
principal bibundles is associative.

\subsection*{Inverse bibundle}
\noindent \\
The inverse bibundle $E^{-1}$ of $E$  has by definition the 
same total space and base space of $E$ but the left action and the right 
actions $\displaystyle \triangleleft^{\!\!-1}$
are defined by 
\eq
h_{\,} 
e^{-1}=(e\triangleleft h^{-1})^{-1} ~~~,~~~ 
e^{-1}\triangleleft^{\!\!-1} h=(h^{-1} e)^{-1} 
\label{action-1}
 \en
here $e^{-1}$ and $e$ 
are the same point of the total space, we write
$e^{-1}$ when the total space is endowed with the $E^{-1}$ principal 
bibundle structure, we write $e$ when the total space is endowed with the
$E$ principal bibundle structure. From the Definition
$(\ref{action-1})$ it follows that 
$h_{}e^{-1}=e^{-1}\triangleleft^{\!\!-1} \vvphi_e(h)$.
Given the sections $\tt^i\,:\,U^i\rightarrow E$ of $E$ we canonically have 
the sections ${\tt^i}^{-1}\,:\,U^i\rightarrow E^{-1}$ of $E^{-1}$
(here again $\tt^i(x)$ and ${\tt^i}^{-1}(x)$ are the same point of the total
space). The section ${\tt^i}^{-1}\tt^i$ of $E^{-1}E$ is central,
i.e. it satisfies (\ref{left&rightiso}). We also have 
${\tt^i}^{-1}\tt^i={\tt^j}^{-1}\tt^j$ in $U^{ij}$; we 
can thus define a canonical (natural) global central section $\Ii$ 
of $E^{-1}E$, 
thus showing that $E^{-1}E$ is canonically trivial.
Explicitly we have $\overline{I}[e'^{-1}, e] = h$ with $e'\triangleleft h = e.$
Similarly for $EE^{-1}$.
The space of isomorphism classes of $H$-bibundles on $M$
[cf. (\ref{isoal})] can now be endowed with a group structure. 
The unit is the isomorphism class of the trivial product bundle  
$M\times H$. The inverse of the class represented by $E$ is the 
class represented by $E^{-1}$.

\sk
Consider two isomorphic bibundles $E$ and $E'$ on $M$. The choice of a
specific isomorphism between $E$ and $E'$ is equivalent to the choice of
a global central section of the bibundle $EE'^{-1}$, i.e. a global section 
that satisfies (\ref{left&rightiso}).
Indeed, let $\ff$ be a global section of $EE'^{-1}$, given 
an element $e\in E$ with base point $x\in M$, there is a unique
element  $e'^{-1}\in E'^{-1}$ with base point $x\in M$
such that $[e,e'^{-1}]=\ff(x)$. Then the isomorphism $E\sim E'$
is given by $e\mapsto e'$; it is trivially compatible with the 
right $H$-action, it is compatible with the left $H$-action because of 
the centrality of $\ff$.

More generally let us consider two isomorphic left $H$-bundles $E
\stackrel{W}{\sim} E'$ which are not necessarily bibundles. 
Let us write a generic element $(e,e')\in E\oplus E'$ in the form
$(e, hW(e))$ with a properly chosen $h\in H$. We introduce an equivalence
relation on $E\oplus E'$ by $(e, hW(e))\sim (h'e, hW(h'e))$. Then 
$T=E\oplus E'/\sim$ is a trivial left $H$-bundle with
global section $\bar{\sigma}([e, hW (e)])=h^{-1}$
(the left $H$-action
is inherited from $E$). Recalling the
comments after Proposition $4$, we equip $T$ with
trivial $H$-bibundle structure and global central section $\bar{\sigma}$.
Next we consider the product $TE'$
and observe that any
element $[[e,e'_1],e''_2]\in TE'$ can be written as 
$[[\tilde e, W(\tilde e)],W(\tilde e)]$ with a unique $\tilde e \in
E$. We thus have a canonical isomorphism between $E$ and $TE'$ and
therefore we write $E=TE'$. Vice versa if $T$ is a
trivial bibundle with  global central section $\bar\sigma : T \rightarrow H$
and $E$,$E'$ are left $H$-bundles and  $E=TE'$, i.e $E$ is canonically 
isomorphic to $TE'$, then we can consider the isomorphism 
$E \stackrel{W}{\sim} E'$ defined by $W([t, e'])=\bar\sigma (t)e'$
(here $[t,e']$ is thought as an element of $E$ because of the identification
$E=TE'$). It is then easy to see that the trivial bibundle with
section given by this isomorphism $W$ is canonically isomorphic 
to the initial bibundle $T$.

We thus conclude that the choice of an isomorphism
between two left $H$-bundles $E$ and $E'$ is
equivalent to the choice of a trivialization  
(the choice of a global central section) of the bibundle $T$,
in formulae
\eq
E \stackrel{W}{\sim} E'~~~~~~\Longleftrightarrow E=TE'
\en
where $T$ has a given global central section. 

\subsection*{Local coordinates description}
\noindent \\
We recall that an atlas of charts for an $H$ principal left bundle
$E$ with base space $M$ is given by a covering $\{U^i\}$ of $M$, 
together with sections $\tt^i\,:\,U^i\rightarrow E$ (the sections $\tt^i$
determine isomorphisms between the restrictions of $E$ to $U^i$
and the trivial bundles $U^i\times H$). The transition functions 
$h^{ij}~:~U^{ij}\rightarrow H$ are defined by $\tt^i=h^{ij}\tt^j$. 
They satisfy on $U^{ijk}$ the cocycle condition
$$
h^{ij}h^{jk}=h^{ik}\,.
$$
On $U^{ij}$ we have $h^{ij}={h^{ji}}^{-1}$.
A section 
$\ss~:~U\rightarrow E$ has local representatives $\{ s^i\}$ where 
$s^i~:~U\cap U^i\rightarrow H$ and in $U^{ij}$ we have 
\eq
s^ih^{ij}=s^j~.
\en
If $E$ is also a bibundle we set
\eq
\vphi^i\equiv \vvphi_{\ttt^i}~:~U^i\rightarrow Aut(H) \label{vphimap}
\en
and we then have
$
\,\forall ~h\in H~,
~~\vphi^i(h)h^{ij}=h^{ij}{\vphi^j}(h)\;,\;
$ i.e.
\eq
Ad_{h^{ij}}=\vphi^i\circ {\vphi^j}^{-1} \label{hvphi}~.
\en
We call the set $\{h^{ij},\vphi^i\}$ of transition functions and
$\vphi^i$ maps satisfying (\ref{hvphi}) a set of local data of $E$.
A different atlas of $E$, i.e. a different choice of sections $\tt'^i=r^i\tt^i$
where $r^i\,:\,U^i\rightarrow H$ (we can always refine the two atlases
and  thus choose a common covering
$\{U^i\}$ of $M$), gives local data
\eqa
&&h'^{ij}={r^i}h^{ij}{r^j}^{-1}~\label{equivalence}~,\\
&&\vphi'^i=Ad_{r^i}\circ \vphi^i~\label{phiimaps}~.
\ena
We thus define two sets of local data $\{h^{ij},\vphi^i\}$ and 
$\{h^{ij},\vphi^i\}$ to be equivalent if they are related by
(\ref{equivalence}), (\ref{phiimaps}).

One can reconstruct an $H$-bibundle $E$ from a given set of 
local data $\{h^{ij},\vphi^i\}$ relative to a covering $\{U^i\}$ of $M$. 
{}For short we write 
$E=\{h^{ij},\vphi^i\}$.
The total space of this bundle is the set of triples $(x,h,i)$
where $x\in U^i$, $h\in H$, modulo the equivalence relation 
$(x,h,i)\sim (x',h',j)$ iff $x=x'$ and $ h h^{ij}=h'$. 
We denote the equivalence class by $[x,h,i]$. 
The left $H$ action is 
$h'[x,h,i]=[x,h'h,i]$. The right action, given by 
$[x,h,i]\triangleleft h'=[x,h\vphi^i(h'),i]$ 
is well defined because of (\ref{hvphi}). 
The $h^{ij}$'s are transition functions of
the atlas given by the sections $\tt^i_{\,}:_{\,}U^i\rightarrow E$,
$\,\tt^i(x)=[x,1,i]$, and we have $\vvphi_{\ttt^i}=\vphi^i$.
It is now not difficult to prove that
equivalence classes of local data are in one-to-one correspondence
with isomorphism classes
of bibundles. [Hint: ${\tt'^i}^{-1}(r^i\tt^i)$ is central and $i$ independent].
\sk

Given two $H$  bibundles $E=\{h^{ij},\vphi^i\}$ and 
$\tilde E=\{\tilde{h}^{ij},\tilde{\vphi}^i\}$ on the same base space
$M$, the product bundle 
$E\tilde E$ has transition
functions and left $H$-actions given by
(we can always choose a covering $\{U^i\}$ of $M$ common to $E$ and 
$\tilde{E}$)
\eq
E\tilde E=\{h^{ij}\vphi^j(\tilde{h}^{ij})\,,
\vphi^i\circ\tilde{\vphi}^i\}\label{productbundle}
\en
If $\tilde E$ is not a bibundle the product $E\tilde E$ is still a 
well defined bundle with transition functions 
$h^{ij}\vphi^j(\tilde{h^{ij}})$.
Associativity of the product (\ref{productbundle}) is easily verified.
One also shows that if $s^i\,,\tilde{s}^i\::\;U\cap U^i\rightarrow H$ 
are local representatives for the sections  $\ss~:~U\rightarrow E$ and  
$\tilde{\ss}~:~U\rightarrow \tilde E$ then the 
local representative for the product section 
$\ss\mbox{\boldmath $\tilde s$}~:~U\rightarrow E\tilde E$ 
is given by 
\eq
s^i\vphi^i(\tilde{s}^i)~.
\en 
The inverse bundle of $E=\{h^{ij},\vphi^i\}$ is 
\eq
E^{-1}=\{ {\vphi^j}^{-1} ({h^{ij}}^{-1})\,,{\vphi^i}^{-1} \}
\label{inversebund}
\en
(we also have
$
{\vphi^j}^{-1}({h^{ij}})^{-1}\!={\vphi^i}^{-1}({h^{ij}}^{-1})\,
$). If $\ss\;:\;U\rightarrow E$ is a section of $E$ with
representatives $\{s^i\}$ then 
$\ss^{-1}\;:\;U\rightarrow E^{-1}$, has representatives
$\{{\vphi^i}^{-1}({s^i}^{-1})\}$.

\sk
A trivial bundle $T$ with global central section $\tt$, in an
atlas of charts subordinate to a cover $U^i$ of the base space $M$, 
reads
\eq
T=\{f^i{f^j}^{-1}, \,Ad_{f^i}\}~,
\en 
where the section $\tt\equiv\ff^{-1}$ has local representatives 
$\{{f^i}^{-1}\}~.$
{}For future reference notice that
$\,T^{-1}=\{{f^i}^{-1}{f^j}, \,Ad_{{f^i}^{-1}}\}~$ has global
central  section $\ff=\{f^i\}$, and that $E_{\,}T^{-1}_{\,}E^{-1}$ 
is trivial,
\eq
E_{\,}T^{-1}_{\,}E^{-1}=\{{\vphi^i}({f^i}^{-1}){\vphi^j}({f^j}),\,
Ad_{{\vphi^i}({f^i}^{-1})}\}~.\label{conj}
\en
We denote by $\mbox{\boldmath $\vphi$}(\ff)$ the 
global central section
$\{{\vphi^i}({f^i})\}$  of $E_{\,}T^{-1}_{\,}E^{-1}$. 
Given an {\sl arbitrary} section $\ss\;:\;U\rightarrow E$, we have, in $U$  
\eq
\mbox{\boldmath $\vphi$}(\ff)=\ss_{}\ff_{}\ss^{-1}  \label{globalsect}
\en
{\sl Proof : }
$\ff_{}\ss^{-1}=\{_{}
f^iAd_{{f^i}^{-1}}({\vphi^i}^{-1}({s^i}^{-1}))_{}\}=\{{\vphi^i}^{-1}({s^i}^{-1})f^i\}$
and therefore 
$\ss\ff_{}\ss^{-1}=\{_{}\vphi^i(f^i)_{}\}=\mbox{\boldmath{{${\vphi}$}}}(\ff)$.
Property (\ref{globalsect}) is actually the defining property of
$\vvphi(\ff)$.  Without using an atlas of charts, 
we define the global section $\vvphi(\ff)$ of $ET^{-1}E^{-1}$ to be that
section that locally satisfies  (\ref{globalsect}). The definition is well
given because centrality of $\ff$ implies that
$\mbox{\boldmath $\vphi$}(\ff)$ is independent from $\ss$. 
Centrality of the global section $\ff$ also implies that 
$\mbox{\boldmath $\vphi$}(\ff)$ is a  global central section. If $\overline{\sigma}$ is the global central section of $T$,
the corresponding global section $\overline{\sigma}'$of $E_{\,}T^{-1}_{\,}E^{-1}$ is $\overline{\sigma}'[e,t,e'^{-1}]=
\vphi_e(\overline{\sigma}(t)) h$ with $e=he'$.

\sk
The pull-back of a bi-principal bundle is again a  bi-principal bundle.
It is also easy to verify that the pull-back commutes with the product.

\subsection*{$\Dd$-$\Hh$ bundles}
\noindent\\ 
We can generalize the notion of a bibundle by introducing 
the concept of a crossed module.

We say that $H$ is a crossed $D$-module \cite{Brown} 
if there is a group homomorphism $\alpha: H\rightarrow D$ and an
action of $D$ on $H$ denoted as $(d, h)\mapsto \,^dh$ such that
\eq
 \forall h,\, h'\in H~,~~~~~~~~~\,^{\alpha(h)}h'=hh'h^{-1}\hskip 1cm \label{cross1}
\en
and forall $h\in H$, $d\in D$,
\eq
\alpha (\,^d h)= d \alpha(h) d^{-1} \label{cross2}
\en
holds true. 

Notice in particular that $\al(H)$ is normal in $D$. 
The canonical homomorphism 
$Ad: H\rightarrow Aut(H)$ and the canonical action of
$Aut(H)$ on $H$ define on $H$ the structure
of a crossed $Aut(H)$-module.
Given a $D$-bundle $Q$ we can use the homomorphism 
$t: D \rightarrow Aut(H), \,t\circ \alpha=Ad$ to form
$Aut(H) \times_D Q$. 
\begin{definition}
Consider a left $D$-bundle $Q$ on $M$ such that the $Aut(H)$-bundle
$Aut(H) \times_D Q$ is trivial. Let $\sigma$ be a global section of
$Aut(H) \times_D Q$. We call the couple $(Q,\sigma)$ a $D$-$H$ bundle.
\end{definition}
Notice that if $\overline{\sigma}: Aut(H) \times_D Q \ni [\eta,q] \mapsto 
\overline{\sigma}([\eta, q]) \in Aut(H)$ is a global section of 
$Aut(H) \times_D Q$ then
\newline 
i) the automorphism $\vvpsi_q \in Aut(H)$ defined by
\eq
\vvpsi_q \equiv \overline{\sigma}([id, q])
\en
is  $D$-equivariant, 
\eq
\vvpsi_{dq}(h)=~^d_{\!}\vvpsi_q(h)~\label{triv}
\en
\newline
ii) the homomorphism $\vvxi_q:H \rightarrow D$ defined by 
\eq
\vvxi_q(h) \equiv \alpha\circ\vvpsi_q(h) \label{alphacircpsi}
\en 
gives a fiber preserving action $q\triangleleft h
\equiv \vvxi_q(h)q$ of $H$ on the right, commuting with the left $D$-action.
i.e. 
\eq
\forall\;h\in H, d\in D , q\in Q,~~~~~~\mbox{ $~~~~$ }
(d \,q)\triangleleft h=d_{}( q\triangleleft h)~\,.
\en 
Vice versa we easily have 
\begin{proposition}
Let $H$ be a crossed $D$-module. If $Q$ is a left $D$ bundle admitting a right
fiber preserving $H$ action commuting with the left $D$ action,
and the homomorphism $\vvxi_q:H \rightarrow D$, defined by
$q\triangleleft h = \vvxi_q(h)q$ is of the form (\ref{alphacircpsi})
with a $D$-equivariant $\vvpsi_q \in 
Aut(H)$ [cf. (\ref{triv})], then $Q$ is a $D$-$H$ bundle.
\end{proposition} 
There is an obvious notion of an isomorphism between two $D$-$H$ bundles
$(Q,\sigma)$ and $(\tilde Q, \tilde{\sigma})$; it is an isomorphism between $D$-bundles $Q$ and 
$\tilde Q$ intertwining between $\sigma$ and $\tilde{\sigma}$ .  
In the following we denote a $D$-$H$ bundle $(Q, \sigma)$ simply as
$Q$
without spelling out explicitly the choice of a global section
$\sigma$ of $Aut(H)\times_D Q$.
As in the
previous section out of a given isomorphism we can construct
a trivial $D$-bibundle $Z$ with a global central section $\zz^{-1}$ 
such that $\tilde Q$ and $ZQ$ are canonically identified and we again 
write this as $\tilde Q = ZQ$. The $\vvpsi$ map of $Z$ is given by
$Ad_{\overline z^{-1}}$. 

Note that the product of a trivial $D$-bibundle 
$Z$ and a $D$-$H$ bundle $Q$ is well-defined and 
gives again a $D$-$H$ bundle.

The trivial bundle $M\times D \rightarrow M$, with 
right $H$-action given by $(x,d)\triangleleft h= (x,d\alpha(h))$,
is a 
$D$-$H$ bundle, we have $\vvpsi_{(x,d)}(h)=~^d_{\!}h$.
A $D$-$H$ bundle $Q$ is trivial if it is isomorphic to $M\times D$.
Similarly to the case of a bibundle we have that a $D$-$H$ bundle is 
trivial iff it has a global section $\ssigma$ which is central 
with respect to the left and the right actions of $H$ on $Q$,
\eq
\ssigma(x) \triangleleft h = \alpha (h) \ssigma(x)~.
\en
The corresponding map $\overline{\sigma} : Q\rightarrow D$ 
is then bi-equivariant
\eq
\overline{\sigma}(d q \triangleleft h)= d_{\,}\overline{\sigma}(q)\alpha(h).
\en   
The pull-back of a $D$-$H$ bundle is again a $D$-$H$ bundle. 

The trivial bundle $Aut(H) \times_H E$ (cf. Proposition $4$) 
is an $Aut(H)$-$H$ bundle. 
{\sl Proof.} The left $Aut(H)$ and the right $H$ actions commute, 
and they are related by 
$ [\eta, e]h = Ad_{\eta(\vvph_e(h))} [\eta ,e]$; we thus have $\vvpsi_{[\eta,
e]}=\eta\circ \vvphi_e$, which structures $Aut(H)\times_H E$ into an 
$Aut(H)$-$H$ bundle. 
Moreover $\overline{\sigma}([\eta, e]) = \eta\circ\vvphi_e$ 
is bi-equivariant, hence $Aut(H)\times_H E$ is isomorphic to 
$M\times Aut(H)$ as an $Aut(H)$-$H$ bundle. 

More generally, we can use the left $H$-action on $D$ given by the
homomorphism $\alpha :H \rightarrow D$ to associate to
a bibundle $E$ the bundle $D\times_H E$. 
The $H$-automorphism $\vvpsi_{[d,e]}$ defined by $\vvpsi_{[d,e]}=
~^d_{\!}\vvphi_e(h)$ endows $D\times_H E$ with 
a $D$-$H$ bundle structure.

There is the following canonical construction associated with a $D$-$H$
 module. We use the $D$-action on $H$ to form the associated
bundle $H\times_D Q$. 
Using the equivariance property (\ref{triv}) of $\vvpsi_q$ we easily get the
following proposition.
\begin{proposition}\label{D-H-triv}
The associated bundle $H\times_D
Q$ is a trivial $H$-bibundle with actions $h'[h,
q]=[\vvpsi_q(h')h,q]$ and $[h,q]\triangleleft h'=[h\vvpsi_q(h'),q]$,
and with 
global central section given by $\bar\sigma([h,q])= \vvpsi_q^{-1}(h)$.
\end{proposition}

The local coordinate description of a $D$-$H$ bundle $Q$ 
is similar to that of a
bibundle. We thus omit  the details.
We denote by
$d^{ij}$ the transition functions of the left principal $D$-bundle $Q$. Instead of local maps (\ref{vphimap}) 
we now have local maps $\psi^i: U^i \rightarrow Aut(H)$, such that
(compare to (\ref{hvphi}))
\eq
\,^{d_{ij}}h=\psi^i\circ \psi^{j^{-1}}(h). 
\en 
The product $QE$ of a $D$-$H$
bundle $Q$ with a $H$-bibundle $E$ can be defined as in (\ref{12}),
(\ref{13}). 
The result is again a $D$-$H$ bundle. If $Q$ is locally given by $\{d^{ij}, \psi^i\}$ and
$H$ is locally given by $\{h^{ij}, \vphi^i\}$ then $QE$ is locally given by
$\{d^{ij}\xi^j(h^{ij}), \psi^i\circ\vphi^i\}$. Moreover if $Z=\{z^iz^{j^{-1}},
Ad_{z^i}\}$ 
is a trivial $D$-bibundle with section $\zz^{-1}=\{{z^i}^{-1}\}$, then 
the well-defined $D$-$H$ bundle $ZQ$ is locally given by
$\{z^i d^{ij}z^{j^{-1}}, \,^{z^{i}}\circ \psi^i\}$.
 He have the following
associativity property
\eq
(ZQ)E = Z(QE)~,
\en
and the above products commute with pull-backs.

Given a $D$-$H$ bundle $Q$ and a trivial
$H$-bibundle $T$ with section $\ff^{-1}$ there exists a unique 
trivial $D$-bibundle 
$\vvxi(T)$ with section $\vvxi(\ff^{-1})$ such that
\eq
Q T = \vvxi(T) Q~, \label{ximap}
\en
i.e. such that for any local section $\ss$ of $Q$  one has 
$\ss\ff^{-1}=\vvxi(\ff^{-1})\ss$. The notations $\vvxi(T)$,
$\vvxi(\ff^{-1})$ are inferred from the local expressions of 
these formulae.
Indeed, if locally $T = \{{f}^i{f}^{j^{-1}},
Ad_{{f}^i}\}$ and $\ff=\{f^i\}$, then
$\vvxi(T)=$ $\{\xi^i({f}^i) \xi^j({f}^{j})^{-1},
Ad_{\xi^i(f^i)}\}$ and $\vvxi(\ff)=\{\xi^i(f^i)\}$.

{}Finally, as was the case for bibundles, 
we can reconstruct a $D$-$H$ bundle $Q$ from a given set of local data 
$\{d^{ij}, \psi^i\}$ relative to a covering $\{U^i\}$ of $M$. 
Equivalence of local data for $D$-$H$ bundles is defined in
such a way that isomorphic (equivalent) $D$-$H$ bundles have 
equivalent local data, and vice versa.
\sk
\sk

\sk
\section{\bf Connection and Curvature on Principal Bibundles}

Since a bibundle $E$ on $M$ is a bundle on $M$ that
is both a left principal $H$-bundle and a right principal $H$-bundle,
one could then define a connection on a bibundle to be a 
one-form $\ax$ on $E$ that is both a left and a right principal
$H$-bundle connection.
This definition [more precisely the requirement ${\mathcal A}^r=0$ in
 (\ref{requiv})] preserves the left-right
symmetry property of the bibundle structure, but it turns out to be
too restrictive, indeed not always a bibundle can be endowed with such
a connection, and furthermore the
corresponding curvature is valued in the center of $H$.
If we insist in  preserving the
left-right symmetry structure we are thus led to
generalize (relax) 
the notion of connection.
In this section we will
see that a connection on a bibundle is a couple $(\ax, A)$ where
$\ax$ is a one-form on $E$ with values in ${\mlie}(H)$
while  $A$ is a ${\mlie}(Aut(H))$ valued one-form on $M$.
In particular we see that if $A=0$ then $\ax$  is a left connection on
$E$ where $E$ is considered just as a left principal bundle.
We recall that a connection $\ax$ on a left principal bundle $E$
satisfies \cite{Koba-Nomizu}
\sk
\noi{\it{i})} the
pull-back of $\ax$ on the fibers of $E$ is the 
  right invariant Maurer-Cartan one-form. Explicitly, let $e\in E$, let $g(t)$ be a
  curve from some open interval $(-\varepsilon,\varepsilon)$ of the real line
  into the group $H$ with $g(0)=1_H$, and let $[g(t)]$ denote the
  corresponding tangent vector in $1_H$  and $[g(t)e]$ the vertical
  vector based in $e\in E$. Then 
\eq
\ax[g(t)e]=-[g(t)]~. \label{lcm}
\en
Equivalently $\ax[g(t)e]=\zeta_{[g(t)]}$ where $\zeta_{[g(t)]}$ 
is the right-invariant vector field associated with ${[g(t)]}\in
{\mlie}(H)$, i.e. $\zeta_{[g(t)]}|_{_{h}}=-{[g(t)h]}$.
\sk
\noi
{\it{ii})} under the left $H$-action we have the 
equivariance property
\eq
{l^h}^*\ax=Ad_h\ax\label{leftconn}
\en
where $l^h$ denotes left multiplication by $h\in H$.
\sk
\noi Now property {\it{i})} is compatible with  the bibundle structure on
$E$ in the following sense, if $\ax$ satisfies {\it{i})} then 
$-\vvphi^{-1}(\ax)$ pulled back on the fibers is the left invariant 
Maurer-Cartan one-form
\eq
-\vvphi^{-1}(\ax)[e g(t)]=[g(t)] \label{rcm}~,
\en
here with abuse of notation we use the same symbol $\vvphi^{-1}$
for the map $\vvphi^{-1}\,:E\times H\rightarrow H$ and its
differential map $\vvphi_*^{-1}\,:E\times {\mlie}(H)\rightarrow
{\mlie}(H)$. Property (\ref{rcm}) is equivalent to 
$\ax[g(t)e]=\xi_{[g(t)]}$ where $\xi_{[g(t)]}$ 
is the left-invariant vectorfield associated with ${[g(t)]}\in
{\mlie}(H)$, i.e. $\xi_{[g(t)]}|_{_{h}}={[hg(t)]}$.
Property (\ref{rcm}) is easily proven, $$-\vvphi^{-1}(\ax)[e
g(t)]=-\vvphi_e^{-1}(\ax[\vvphi_e(g(t))e])=
\vvphi^{-1}_e[\vvphi_e(g(t))]=[g(t)]~.$$
Similarly, on the vertical vectors $v_{_V}$ of $E$ we have
$\left({r^h}^*\ax-\ax\right)(v_{_V})=0~,$ 
$\left({l^h}^*\vphi^{-1}(\ax)- \right.$
$\left.\vphi^{-1}(\ax)\right)(v_{_V})$ $=0\,$ and 
\eqa
&&\left({l^h}^*\ax-Ad_h\ax\right)(v_{_V})=0~~~,\label{twovv}\\
&&\left({r^h}^*\vphi^{-1}(\ax)-Ad_{h^{-1}}\vphi^{-1}(\ax)\right)(v_{_V})=0~~~.
\label{threevv}
\ena
On the other hand property {\it{ii})} is not compatible with the
bibundle structure, indeed if $\ax$ satisfies (\ref{leftconn}) then 
it can be shown (see later) that $-\vvphi^{-1}(\ax)$ satisfies
\eq
{r^h}^*\vvphi^{-1}(\ax)=Ad_{h^{-1}}\vvphi^{-1}(\ax)-p^*T'(h^{-1})
\label{connectionr}
\en
where $T'(h)$ is a given one-form on the base space $M$, and
$p\,:\,E\rightarrow M$.
In order to preserve the
left-right symmetry structure we are thus led to
generalize (relax) the equivariance property {\it{ii})} of a connection. 
Accordingly with (\ref{twovv}) and (\ref{connectionr}) we thus require 
\eq
{l^h}^*\ax=Ad_h\ax+ p^*T(h)\label{lequiv111}
\en
where $T(h)$ is a one-form on $M$.
From (\ref{lequiv111}) it follows 
\eq
T(hk)=T(h)+Ad_h T(k)~,\label{1coc}
\en
i.e., $T$ is a 1-cocycle 
in the group cohomology of $H$ with values in ${\mlie}(H)\otimes \Omega^1(M)$. 
Of course if $T$ is a coboundary, i.e. $T(h)=h\chibold h^{-1} -\chibold$ with
$\chibold\in {\mlie}(H)\otimes \Omega^1(M)$, then $\ax+\chibold$ is a
connection. We thus see that eq. (\ref{lequiv111}) is a nontrivial
generalization of the equivariance property only if
the cohomology class of $T$ is nontrivial. 
\sk
Given an element  $X\in{\mlie}(Aut(H))$, we can construct a
corresponding 1-cocycle $T_X$ in the following way,
$$
T_X(h) \equiv[ h e^{tX}(h^{-1})]~,
$$
where $[ h e^{tX}(h^{-1})]$ is the tangent vector to the curve 
$h e^{tX}(h^{-1})$ at the point $1_H$; if $H$ is normal in $Aut(H)$ 
then $e^{tX}(h^{-1})=e^{tX}h^{-1}e^{-tX}$ and we simply have 
$T_X(h)=hXh^{-1}-X$. Given a ${\mlie}(Aut(H))$-valued one-form $A$ on $M$,
we write $A=A^\rho X^\rho$ where $\{X^\rho\}$ is a basis of
${\mlie}(Aut(H))$. We then define $T_A$ as
\eq
T_A\equiv A^\rho T_{X^\rho}~.
\en
Obviously, $p^*T_A=T_{p^{*\!}A}$. Following these considerations
we define
\begin{definition}
A $2$-connection on $E$ is a couple $(\ax, A)$ where: 

\noi i{$)$}$~\ax$ is a
${\mlie}(H)$ valued one-form on $E$ 
such that its pull-back on the fibers of $E$ is the right invariant
Maurer-Cartan one-form, i.e. $\ax$ satisfies (\ref{lcm}),

\noi ii{$)$} $~A$ is a ${\mlie}(Aut(H))$ valued one-form on $M$, 

\noi iii{$)$}$~$ the couple $(\ax, A)$  satisfies
\eq
{l^h}^*\ax=Ad_h\ax+ p^*T_A(h)\label{lequiv}~.
\en
\end{definition}

This definition seems to break the
left-right bibundle symmetry since, for example, only the left $H$
action has been used. This is indeed {\sl not} the case
 
\begin{theorem}\label{r2c}
If $(\ax,A)$ is a $2$-connection on $E$ then 
$(\ax^r, A^r)$, where $\ax^r\equiv-\vvphi^{-1}(\ax)$,
satisfies (\ref{lcm}) and (\ref{lequiv}) with the left $H$ action
replaced by the right $H$ action (and right-invariant vectorfields
replaced by left-invariant vectorfields), 
i.e. it satisfies (\ref{rcm}) and
\eq
{r^h}^*\ax^r=Ad_{h^{-1}}\ax^r+ p^*T_{A^r}(h^{-1})\label{requiv}~,
\en
here $A^r$ is the one-form on $M$ uniquely defined by the property 
\eq
p^{*\!}A^r=\vvphi^{-1}(p^{*\!}A+ad_{\axs})\vvphi+\vvphi^{-1}d\vvphi~.
\en
\end{theorem}
\begin{proof} 
First we observe that from (\ref{lcm}) and (\ref{lequiv}) we have
\eq
{l^{h'}}^*\ax=Ad_{h'}\ax+ p^*T_A(h')+h'dh'^{-1}
\en
where now $h'=h'(e)$, i.e. $h'$ is an $H$-valued function on the total
space $E$. Setting $h'=\vvphi(h),$ with $h\in H$ we have 
\eqa
{r^h}^*\ax={l^{\vvph(h)}}^*\ax&=&Ad_{\vvph(h)}\ax
+ p^*T_A(\vvphi(h))+\vvphi(h)d\vvphi(h^{-1})\nonumber\\
&=&\ax+{\vvphi}(T_{{\mathcal A}^r}(h))~\label{rha}
\ena
in equality (\ref{rha}) we have defined
\eq
{\mathcal A}^r\equiv
\vvphi^{-1}(p^{*\!}A+ad_\axs)\vvphi+\vvphi^{-1}d\vvphi\,. \label{Ar}
\en
Equality
(\ref{rha}) holds
because  of  
the following properties of $T$,
\eqa
&&T_{\vvph^{-1}d\vvph}(h)={\vvphi^{-1}}\left(\vvphi(h)d\vvphi(h^{-1})\right)~,
\label{Tpone}\\
&&T_{\vvph^{-1}p^{*\!}A\vvph}(h)={\vvphi^{-1}} 
\left( T_{p^{*\!}A_{}}(\vvphi(h))^{}_{}\right)~,\label{Tptwo}\\
&&T_{ad_\axs}(h)=Ad_h\ax-\ax~.\label{Tpthree}
\ena
{}From (\ref{rha}), applying $\vvphi^{-1}$ and then using (\ref{extra}) 
one obtains
\eq
{r^h}^*\ax^r=Ad_{h^{-1}}\ax^r
+T_{{\mathcal A}^r}(h^{-1})~.
\label{rhvaa}
\en
Finally, comparing (\ref{threevv}) with (\ref{rhvaa}) we deduce that 
for all $h\in H$, $T_{{\mathcal A}^r}(h)(v_{_V})=0$, and this relation 
is equivalent to ${\mathcal A}^r(v_{_V})=0$. 
In order to prove that ${\mathcal A}^r=p^{*\!}A^r$ where $A^r$ is a
one-form on $M$, we then just need to show that ${\mathcal A}^r$ 
is invariant under the $H$
action, ${l^h}^*{\mathcal A}^r={\mathcal A}^r$. This is indeed the case because
${l^h}^*(\vvphi^{-1}d\vvphi)=\vvphi^{-1}Ad_{h^{-1}} d_{}Ad_{h}\vvphi
=\vvphi^{-1}d\vvphi$, and because
\eqa
{l^h}^*\left(\vvphi^{-1}(p^{*\!}A+ad_\axs)\vvphi\right)
&=&\vvphi^{-1}Ad_{h^{-1}}(p^{*\!}A+{l^h}^* ad_\axs)
Ad_{h}\vvphi\nonumber\\
&=&\vvphi^{-1}\left(Ad_{h^{-1}}p^{*\!}A Ad_{h}+ ad_\axs
  +ad_{Ad_{h^{-1}}T_{p^{*\!}A}(h)}\right)\vvphi
\nonumber\\
&=&
\vvphi^{-1}(p^{*\!}A+ad_\axs)\vvphi\nonumber~.
\ena
\end{proof}
Notice that if $(\ax, A)$ and $(\axp, A')$ are $2$-connections on $E$
then so is the affine sum 
\eq
{\big(}\,p^{*}(\lambda)\ax+(1-p^*\lambda)\axp\,,\,\lambda A+(1-\lambda)A'\,\big{)}
\en
for any (smooth) function $\lambda$ on $M$. 
\sk
As in the case of principal bundles we define a vector $v\in T_eE$ to be
horizontal if $\ax(v)=0$. The tangent space $T_eE$ is then decomposed 
in the direct sum of its horizontal and vertical subspaces; for all 
$v\in T_eE$, we write $v=\H v+\V v$, 
where $Vv=[e^{-t\axs(v)}e]$. 
The space of horizontal vectors 
is however not invariant under the usual left $H$-action, indeed 
$$\ax({l^h}_*(\H v))
=T_A(h)(v)~,$$ 
in this formula, as well as in the
sequel, with abuse of notation $T_A$  stands for $T_{p^{*\!}A}$.
\begin{remark}
It is possible to construct a new left $H$-action ${\LL}_*$ 
on $T_*E$, 
that is compatible with the direct sum decomposition 
$T_*E=\H T_*E+\V T_*E$. We first define, 
for all $h\in H$,
\eqa
L^h_A~:~T_*E&\rightarrow& \V T_*E\nonumber ~,\\
T_eE\ni v &\mapsto&  [e^{tT_A(h)(v)}he]\in \V T_{he}E~,
\ena
and notice that $L_A^h$ on vertical vectors is zero, therefore
$L_A^h\circs L_A^h=0$. We then consider the tangent space map,
\eq
{\LL}^h_*\equiv l^h_*+L^h_A~.
\en
It is easy to see that
${\LL}^{hk}_*={\LL}^h_*
\circs {\LL}^k_*$
and therefore that  ${\LL}_*$ 
defines an action of $H$ on $T_*H$.
We also have 
\eq
{\LL^h}^*\ax=Ad_h\ax~.
\en
{}Finally the action $\LL^h_*$
preserves the horizontal and vertical decomposition 
$T_*E=\H T_*E+\V T_*E$, indeed
\eq
\H \LL^h_*v=\LL^h_*\H v~~~~~~~~~~~~~\mbox{$~$}~~~~~~,~
\mbox{$~$}~~~~~~~~~~~~~~~~~~~~~~~~~~~\V \LL^h_*v=\LL^h_*\V v
\en
{\it Proof.} Let $v=[\gamma(t)]$. Then 
$\H\LL^h_*v=\H l^h_*v= [h\gamma(t)]-[e^{-t\axs[h\gamma(t)]}e]=
[h\gamma(t)]+[e^{t({l^h}^*\axs)(v)}he]=
[h\gamma(t)]+[he^{t\axs(v)}e]+[e^{tT_A(h)(v)}he]=
\LL_*^h(v+[e^{t\axs(v)}e])=\LL^h_*\H v~.$
\end{remark}
\sk
\noi{\bf Curvature.}

\noi 
An $n$-form $\varthee$ is said to be horizontal if $\varthee(u_1,u_2,\ldots
u_n)=0$ whenever 
at least one of the vectors 
$u_i\in T_eE$ is vertical.
The exterior covariant derivative $D\omm$ of an $n$-form $\omm$ 
is the $(n+1)$-horizontal form defined by 
\eq
D\omm(v_1,v_2,\ldots,v_{n+1})
\equiv d\omm(\H v_1,\H v_2,\ldots,\H v_{n+1})
-(-1)^nT_A(\omm)(\H v_1,\H v_2,\ldots,\H v_{n+1})
\en
for all $v_i\in T_eE$ and $e\in E$. In the above formula $T_{A}(\omm)$ is defined by
\eq
T_{A}(\omm)\equiv \omm^\al \wedge {T_{\!A}}_*(X^\al)~,\label{defTAa}
\en
where ${T_{\!A}}_*\,:\,{\mlie}(H)\rightarrow {\mlie}(H)\otimes
\Omega^1(E)$ is the differential of 
$T_{A}\,:\,H\rightarrow {\mlie}(H)\otimes \Omega^1(E)$. 
If $H$ is normal in $Aut(H)$ we simply have 
$T_A(\omm)
= \omm^\rho \wedge p^{*\!}A^\sigma[X^\rho,X^\sigma]=[\omm,p^{*\!}A]$,
where now $X^\rho$ are generators of ${\mlie}(Aut(H))$.

\sk

The $2$-curvature of the $2$-connection  $(\ax,A)$
is given by the couple
\eq
(\kx,K)\equiv \big( D\ax\,,\, dA+A\wedge A \big).
\en
We have the Cartan structural equation
\eq
\kx=d\ax + {1\over 2}[\ax, \ax] + T_{A}(\ax) ~,
\label{CSE} 
\en
where $\frac{1}{2}[\ax, \ax]=\frac{1}
  {2}\ax^\al\wedge\ax^\be[X^\al,X^\be]
=\ax\wedge\ax$ with 
$X^\al\in {\mlie}(H)$,

The proof of eq.~(\ref{CSE}) is very similar to the usual proof of the
Cartan structural equation for principal bundles.
One has just to notice that the extra term $T_A(\ax)$ is necessary
since  $d\ax(\V v, \H u)=-\ax([\V v, \H u])={T_{\!A}}_*(\ax(\V v))(\H
u)=-T_A(\ax)(\V v, \H u)$.

The $2$-curvature $(\kx,K)$ satisfies the following
generalized equivariance property
\eq
{l^{h}}^*\kx = Ad_h\kx  + T_K(h)~,
\en
where with abuse of notation we have written $T_K(h)$ instead of
$T_{p^*K}(h)$.
We also have the Bianchi identities, $dK+A\wedge K=0$ and
\eq 
\label{Bianchi}
D\kx=0
~.
\en

Given an horizontal $n$-form $\varthee$ on $E$ that is
$\Theta$-equivariant, i.e. that satisfies 
${l^{h}}^*\varthee = Ad_h\varthee + T_\Theta(h)~,$
where $\Theta$ is an $n$-form on $M$, we have the structural equation
\eq
D\varthee=d\varthee+[\ax,\varthee] + T_\Theta(\ax)-(-1)^nT_A(\varthee)
\label{SEqua}~~
\en
where 
$[\ax,\varthee]=
\ax^\al\wedge\varthee^\be[X^\al,X^\be]=\ax\wedge\varthee -(-1)^n \varthee\wedge\ax$.
The proof is again similar to the usual one (where $\Theta=0$) and is 
left to the reader. We also have that $D\varthee$ is 
$(d\Theta+[A,\Theta])$-equivariant,
\eq
{l^h}^*D\varthee=Ad_h\varthee+T_{d\Theta+[A,\Theta]}(h)~.
\en
Combining (\ref{SEqua}) and (\ref{Bianchi})
we obtain the explicit expression of the Bianchi identity
\eq
d \kx +[\ax, \kx]
+T_{K}(\ax) -T_{A}(\kx)=0
\en
We also have
\eq \label{Dsquare}
D^2\varthee=[\kx,\varthee]+T_\Theta (\kx)-(-1)^nT_K(\varthee)~.
\en
\sk

As was the case for the $2$-connection $(\ax,A)$, also for the  
$2$-curvature $(\kx,K)$ we can have a formulation using the right $H$
action instead of the left one. Indeed one can prove 
that if $(\kx,K)$ is a $2$-curvature then $(\kx^r,K^r)$ where
$$
\kx^r=-\vvphi^{-1}(\kx)~~~,~~~~~K^r=\vvphi^{-1}(K+ad_\kxs)\vvphi
$$ 
is the right $2$-curvature associated with the right $2$-connection  
$(\ax^r,A^r)$. In other words we have that
$\kx^r$ is horizontal and that
$$
\kx^r = \kx_{\axs^r}~~~,~~~~K^r=K_{A^r}
$$
(for the proof we  used $T_{A^r}(\vvphi^{-1}(X))
=\vvphi^{-1}([X,\ax]+T_A(X))
+d\vvphi^{-1}(X)$, $X\in {\mlie}(H)$).
We also have
\eq
{r^h}^*\kx^r=Ad_{h^{-1}}\kx^r
+T_{K^r}(h^{-1})~.~~
\label{rhva}
\en
More in general consider the couple $(\varthee, \Theta)$  where 
$\varthee$,  is an horizontal $n$-form on $E$
that is $\Theta$-equivariant. Then we have the couple $(\varthee^r,\Theta^r)$
where $\varthee^r=-\vvphi^{-1}(\varthee)$ 
is an horizontal $n$-form on $E$
that is right $\Theta^r$-equivariant,
\eq
{r^h}^*\varthee^r=Ad_{h^{-1}}\varthee^r
+T_{\Theta^r}(h^{-1})~.~~
\en  
with $\Theta^r=\vvphi^{-1}(\Theta+ad_{\varthees})\vvphi$.
\sk

The pull-back of a $2$-connection (or of a horizontal form) 
on a principal $H$-bibundle 
is a $2$-connection (horizontal form) on the pulled
back principal $H$-bibundle, moreover the 
exterior covariant derivative -and in particular the definition of 
$2$-curvature- commutes with the pull-back operation. 

\sk

\noi {\bf Local coordinates description.}

\noi Let's consider 
the sections $\tt^i\,:\,U^i\rightarrow E$ 
subordinate to the covering $\{U^i\}$ of $M$. 
Let $\iota\,:\,H\times U^i\rightarrow p^{-1}(U^i)\subset E$ 
be the local trivialization of $E$ induced by $\tt^i$ according to
$\iota(x,h)=h\tt^i(x)$, where $x\in M$.
We define the one-forms on $U^i\subset M$
\eq
a^i={\tt^i}^*\ax~,
\en
then, the local expression of $\ax$ is
$ha^ih^{-1}+T_A(h)+hdh^{-1}$, more precisely,
\eq
\iota^*(\ax)_{(x,h)}(v_{{M}},v_{H}) 
= ha^i(x)h^{-1}(v_{M})+T_{A(x)}(h)(v_{M})+hdh^{-1}(v_{H})~,
\en
where $v_{M}\,$,$v_{H}$ are respectively tangent vectors of $U^i\subset M$
at $x$, and of $H$ at $h$, and where $-hdh^{-1}$ denotes the
Maurer-Cartan one-form on $H$ evaluated at $h\in H$.
Similarly the local expression for $\kx$ is 
$hk^ih^{-1}+T_K(h)$, where $k^i={\tt^i}^*\kx$.

Using the sections $\{\tt^i\}$ 
we also obtain an explicit expression for $A^r$,
\eq
A^r={\tt^i}^*{\mathcal A}^r=\vphi_i^{-1} 
(A+ {ad}_{a^i}) \vphi_i +\vphi_i^{-1}d\vphi_i~.
\en
Of course in $U^{ij}$ we have 
$ {\tt^i}^*{\mathcal A}^r={\tt^j}^*{\mathcal A}^r$, so that $A^r$ is
defined on all $M$.
In $U^{ij}$ we also have
$ a^i=h^{ij}a^j {h^{ij}}^{-1}+h^{ij}d{h^{ij}}^{-1}+T_A(h^{ij})$ and
$k^i=h^{ij}k^j {h^{ij}}^{-1}+T_K(h^{ij})\,$.

\subsection*{Sum of $2$-connections}
\noindent \\
If the group $H$ is abelian, on the product bundle $E_1E_2$ there is 
the natural
connection $\ax_1\+ \ax_2$ obtained from the connections $\ax_1$ 
and $\ax_2$
on $E_1$ and $E_2$. In this subsection we generalize to the nonabelian
case the sum of connections. 
Consider the following diagram
\eq\label{sumconn1}
\xymatrix
         {
         & E_1 \oplus E_{2}  \ar[d]_{\pi_{1}}
        \ar[dr]^{\pi_{\oplus}}\ar[r]^{\hspace {0.5cm}\pi_{2}}  & E_{2} \\
         & E_{1} & E_{1}E_{2} 
         }
\en
and let 
 $(\ax_1,A_2)$ be a $2$-connection on $E_1$ and 
$(\ax_2, A_2)$ a $2$-connection 
 on $E_2$.
Recalling the definition of the product $E_1E_2$, we see that the 
one-form on $E_1 \oplus E_{2}$
\eq
\pi_1^*\ax_1+\vvphi_1(\pi_2^*\ax_2)
\en
is the pull-back of a one-form on $E_1E_2$ iff, for all 
$v_1\in T_{e_1}E$, $v_2\in T_{e_2}E$ and $h\in H$, 
\eqa
\big(\pi_1^*\ax_1+\vvphi_1(\pi_2^*\ax_2)
&&\!\!\!\!\!\!\!\!\!\!\!\!\!\!\big)_{(e_1,e_2)}(v_1, v_2)
\nonumber \\
&=&\big(\pi_1^*\ax_1+\vvphi_1(\pi_2^*\ax_2)\big)_{(e_1h^{-1},h
  e_2)}(r_*^hv_1,l_*^hv_2)\nonumber\\
& & +\, \big( \pi_1^*\ax_1+\vvphi_1(\pi_2^*\ax_2)\big)_{(e_1h^{-1},h
  e_2)}([e_1h^{-1}(t)], [h(t)e_2] )\nonumber
\ena
where $h(t)$ is an arbitrary curve in $H$ with $h(0)=1_H$. 
Since $\ax_1$ and $\ax_2$ satisfy the Cartan-Maurer condition (\ref{lcm}) 
the last addend vanishes identically and therefore the 
expression is equivalent to 
\eq
\pi_1^*\ax_1+\vvphi_1(\pi_2^*\ax_2)=
{r_{\!}l^h}^*\big(\pi_1^*\ax_1+\vvphi_1(\pi_2^*\ax_2)\big)
\label{leftrightcond}
\en
where 
\eqa
\nonumber
r_{\!}l^h\,:\,E_1\oplus E_2 &\rightarrow & E_1\oplus E_2~,\\
(e_1,e_2)&\mapsto & (e_1h^{-1},he_2)~.
\nonumber
\ena
Now, using (\ref{extra}), and then (\ref{rha}) we have
\eqa
\nonumber
{r_{\!}l^h}^*\big(\pi_1^*\ax_1+
\vvphi_1(\pi_2^*\ax_2)\big)
&=&\pi_1^*{r^{h^{-1}}}^*\ax_1+\vvphi_1 Ad_{h^{-1}}
(\pi_2^*{l^h}^*\ax_2)\nonumber\\
&=&\pi_1^*\ax_1+
\vvphi_1(\pi_2^*\ax_2)\nonumber\\
&& +\, \vvphi_1\big(\pi^*_1T_{A_1^r}(h^{-1})
+\pi_2^*Ad_{h^{-1}}T_{A_2}(h)\big)
\nonumber
\ena
and the last addend vanishes iff
\eq
A_2={A_1}^r~.\label{summcond}
\en
In conclusion, when (\ref{summcond}) holds, there exists 
a one-form on $E_1E_2$, denoted by 
$\ax_1\+\ax_2$, such that
\eq
\pi_\oplus^*(\ax_1\+\ax_2)=
\pi_1^*\ax_1+\vvphi_1(\pi_2^*\ax_2)
\label{summ}
\en
{}From this expression it is easy to see that 
$(\ax_1\+\ax_2 , A_1)$ is a $2$-connection on $E_1E_2$.
We then say that 
$(\ax_1,A_1)$ and $(\ax_2,A_2)$ (or simply that $\ax_1$ and $\ax_2$)
are {\it summable} and we write
\eq
(\ax_1,A_1)\,\+\,(\ax_2,A_2)=(\ax_1\+\ax_2,A_1)~.
\en
Notice that the sum operation $\+$ thus defined is associative
(and noncommutative). In other words, if
$\ax_1$ and $\ax_2$ are summable, and if $\ax_2$ and $\ax_3$
are summable then 
$\ax_{1}\+(\ax_2\+\ax_3) =(\ax_{1}\+\ax_2)\+\ax_3$  and 
$(\ax_{1}\+\ax_{2}\+\ax_3,A_1)$ is a $2$-connection on $E_1E_2E_3$.
\sk
\sk
We also have a summability criterion for the couples 
$(\varthee_1, \Theta_1)$ and $(\varthee_2, \Theta_2)$  where 
$\varthee_i$, $i=1,2$  is an horizontal $n$-form on $E_i$
that is $\Theta_i$-equivariant. 
We have that
$(\varthee_1, \Theta_1)\+(\varthee_2, \Theta_2)=(\varthee_1\+ \varthee_2, \Theta_1)$
where
\eq
\pi^*_\oplus(\varthee_1\+ \varthee_2)=
\pi_1^*\varthee_1+\vvphi_1(\pi_2^*\varthee_2)
\en 
is a well defined  horizontal $\Theta_1$-equivariant 
$n$-form on $E_1E_2$ iff 
\eq
\Theta_2={\Theta_1}^r~.
\en
We have 
\eq
(D_{\axs_1}\varthee_1, D_{A_1}\Theta_1 )\+ (D_{\axs_2}\varthee_2,
D_{A_2})\Theta_2= 
(D_{\axs_1 \+\axs_2}(\varthee_1\+\varthee_2), D_{A_1}\Theta_1),
\label{sumforms} 
\en
with obvious notation: $D_{\axs}\varthee = d\varthee+[\ax,\varthee] +
T_\Theta(\ax)-(-1)^nT_A(\varthee)$ and
$D_A\Theta = d\Theta + [A,\Theta]$.
Also the summability of curvatures is a direct consequence of the
summability of their corresponding connections. 
If $(\ax_1,A_1)\,\+\,(\ax_2,A_2)=(\ax_1\+\ax_2,A_1)~$
then
\eq
(\kx_1,K_1)\,\+\,(\kx_2,K_2)=(\kx_1\+\kx_2,K_1)~,\label{sumcurvs}
\en
and we also have
\eq
\kx_{a_1\+a_2}=\kx_1\+\kx_2~.
\label{lastlast}
\en
Summability is preserved under isomorphism, i.e. if
$\ax_i$ are summable connections on $E_i$ ($i=1,2$) and we have
isomorphisms $\sigma_i : E'_i \rightarrow E_i$, then
$\sigma_i^*(\ax_i)$ are summable and
$\sigma_1^*(\ax_2)+\sigma_2^*(\ax_2) = \sigma_{12}^*(\ax_1\+\ax_2)$,
where we have considered the induced isomorphism
$\sigma_{12}\equiv\sigma_{1}\sigma_{2}\,:\:E'_1E'_2\rightarrow E_1E_2$.
The same property holds for horizontal forms.
\sk
\sk
\sk
\section{\bf Nonabelian Bundle Gerbes}
Now that we have the notion of product of principal bibundles
we can define nonabelian bundle gerbes generalizing the construction
studied by Murray \cite{Murray} (see also Hitchin \cite{Hitchin:1999fh} and
\cite{Chat}) in
the abelian case. 

\sk
Consider a submersion $\wp~:~Y\rightarrow M$ (i.e. a map onto with
differential onto) we can always find a covering $\{{O}_{\al}\}$ of $M$
with local sections $\sigma_{\al}~:~{O}_\al\rightarrow Y$,
i.e. $\wp\circ \sigma_\al=id$. The manifold $Y$ will always be 
equipped with the submersion $\wp~:~Y\rightarrow M$. 
We also consider $Y^{[n]}=Y\times_M Y\times_M Y\ldots\times_M Y$ the n-fold
fiber product of $Y$, i.e. 
$Y^{[n]}\equiv\{(y_1,\ldots y_n)\in Y^n\;|\;\wp(y_1)=\wp(y_2)=
\ldots\wp(y_n)\}$. 

Given a  $H$ principal bibundle $\EE$ over $Y^{[2]}$
we denote by $\EE_{12}=p_{12}^*(\EE)$ 
the $H$ principal bibundle on $Y^{[3]}$ obtained as pull-back 
of $p_{12}~:~Y^{[3]}\rightarrow Y^{[2]}$ ($p_{12}$ is the identity on
its first two arguments);
similarly for $\EE_{13}$ and $\EE_{23}$.
\sk
Consider the quadruple
$(\EE,Y,M,\ff)$ where the $H$ principal bibundle on  $Y^{[3]}$, 
$\,\EE_{12}\EE_{23}\,\EE_{13}^{-1}$ is trivial, and $\ff$ is
a global central section of 
$\,\left( \EE_{12}\EE_{23}\,\EE_{13}^{-1}\right)^{-1}$ 
[i.e. $\ff$ satisfies (\ref{left&rightiso})]. 
Recalling the paragraph after formula (\ref{action-1}) 
we can equivalently say that
$\EE_{12}\EE_{23}$ and $\EE_{13}$ are isomorphic, 
the isomorphism being given by the global central section
$\ff^{-1}$ of 
\eq
\,\TT\equiv\EE_{12}\EE_{23}\,\EE_{13}^{-1}~.
\label{trivialT}
\en
We now consider 
$Y^{[4]}$ and the bundles $\EE_{12},\EE_{23},\EE_{13},\EE_{24},
\EE_{34},\EE_{14}$
on $Y^{[4]}$ 
relative to the projections $p_{12}~:~Y^{[4]}\rightarrow Y^{[2]}$ etc.,
and $\TT_{123}^{-1},\TT_{124}^{-1},\TT_{134}^{-1}$ 
relative to $p_{123}~:~Y^{[4]}\rightarrow Y^{[3]}$ etc..
Since the product of bundles commutes with the pull-back of bundles, 
we then have
\eq
\TT^{-1}_{124} \EE_{12}(\TT^{-1}_{234}\EE_{23}\EE_{34})=
\TT^{-1}_{134} (\TT^{-1}_{123}\EE_{12}\EE_{23})\EE_{34}=\EE_{14}
\label{mu4}
\en
as bundles on $Y^{[4]}$.
The first identity in (\ref{mu4}) is equivalent to
\eq
\TT^{-1}_{124} \,\EE_{12}\TT^{-1}_{234}\EE^{-1}_{12}=\TT^{-1}_{134}\,
\TT^{-1}_{123}
\label{mu4bis}
\en
Let us now consider the global central section ${\ff}$ of 
$\TT^{-1}=\EE_{13}\EE_{23}^{-1}\EE_{12}^{-1}$
and  denote by $\ff_{124}$ ($\ff_{234}$, etc.) the global central section of
$\TT^{-1}_{124}$ ($\TT^{-1}_{234}$, etc.) 
obtained as the pull-back of $\ff$. Consistently with (\ref{mu4bis})
we can require the condition
\eq
\ff_{124}{}{\,} \mbox{\boldmath $\vphi$}_{12}(\ff_{234})
=\ff_{134}{}{\,}\ff_{123}\label{cocycleglobal}~
\en
where, following the notation of (\ref{globalsect}), 
$\mbox{\boldmath$\vphi$}_{12}(\ff_{234})$ is the section of
$\TT^{-1}_{234}$ that in any open $\U\subset Y^{[4]}$ equals 
$\ss_{12}\ff_{234}\ss_{12}^{-1}$ where $\ss_{12}\,:\,\U\rightarrow \EE_{12}$
is any section of $\EE_{12}$,
in particular we can choose $\ss_{12}$ to be
the pull-back of a section $\ss$ of $\EE$. 

\begin{definition} 
A Bundle gerbe ${\GG}$ is the quadruple
$(\EE,Y,M,\ff)$ where the $H$ principal bibundle on  $Y^{[3]}$, 
$\,\EE_{12}\EE_{23}\,\EE_{13}^{-1}$ is trivial and $\ff$ is
a global central section of 
$\,\left( \EE_{12}\EE_{23}\,\EE_{13}^{-1}\right)^{-1}$
that satisfies (\ref{cocycleglobal}).
\end{definition}
Recall that when $H$ has trivial centre then the section $\ff$ of
$\TT^{-1}$ is unique; it then follows that relation (\ref{cocycleglobal}) 
is automatically satisfied because the bundle on the l.h.s. and the bundle
on the r.h.s. of (\ref{mu4bis}) admit just one global central section,
respectively $\ff_{124}{}{\,} \mbox{\boldmath $\vphi$}_{12}(\ff_{234})$
and $\ff_{134}{}{\,}\ff_{123}$. Therefore, if $H$ has trivial centre,
a bundle gerbe ${\GG}$ is simply the triple  
$(\EE,Y,M)$, where $\,\EE_{12}\EE_{23}\,\EE_{13}^{-1}$ is
trivial.
\sk

Consider an $H$ principal bibundle $N$ over $Z$ an let 
${\NN}_1=p^*_1(N)$, ${\NN}_2=p^*_2(N)$, be the pull-back of $N$ 
obtained respectively from $p_1:\,Z^{[2]}\rightarrow Z$
and $p_2:\,Z^{[2]}\rightarrow Z$ ($p_1$ projects on the first component,
$p_2$ on the second). If 
$(\EE,Z,M,\ff)$ is a bundle gerbe also
$\left( {\NN}_1\EE {\NN}_2^{-1} ,Z,M,\vvphi_1{(\ff)}\right)$ is a bundle
gerbe. Here $\vvphi_1(\ff)$ is the canonical global central section of
the bibundle ${\NN}_1\TT^{-1} {\NN}_1^{-1}$ and now $\,{\NN}_1$
is the pull-back of $N$ via $p_1\,:\,Z^{[3]}\rightarrow Z$;
locally $\vvphi_1(\ff)=\ss_1\ff\ss_1^{-1}$ where $\ss_1$ 
is the pull-back of any local section $\ss$ of $N$.
Similarly also $\left( \eta_{}\EE,Z,M,_{}
\ll_{13}^{-1}\ff\vvphi_{12}(\ll_{23})\ll_{12}\right)$ 
is a bundle gerbe if $\eta^{-1}$ is a trivial
bundle on $Z^{[2]}$ with global central section $\ll$
(as usual $\vvphi_{12}(\ll_{23})$ denotes the canonical section of
$\EE_{12} \eta_{23}^{-1}\EE_{12}^{-1}$). 
This observations lead to the following definition \cite{Murray2} 
\begin{definition}
Two bundle gerbes ${\GG}=(\EE,Y,M,\ff)$ and ${\GG'}=(\EE',Y',M,\ff')$ 
are stably isomorphic if there exists a bibundle $\NN$ over 
$Z=Y\times_{\!_{M}\!}Y'$ and a trivial bibundle $\eta^{-1}$ 
over $Z^{[2]}$ with section $\ll$
such that 
\eq
{\NN}_1\:q'^*_{\!\!\!\!\!\!\!}\EE'\: {\NN}_2^{-1}=\eta\:
q^*_{\!\!\!\!\!\!\!}\EE
\label{stably}
\en
and
\eq
\vvphi_1(q'^*\!\ff')=
\ll_{13}^{-1}q^*\!\ff\vvphi_{12}(\ll_{23})\ll_{12}
\label{stably1}
\en
where $q^*_{\!\!\!\!\!\!\!}\EE$ and $q'^*_{\!\!\!\!\!\!\!}\EE'$ 
are the pull-back bundles relative to the projections 
$q\,:\,Z^{[2]}\rightarrow Y^{[2]}$ and  $q'\,:\,Z^{[2]}\rightarrow 
Y'^{[2]}$. Similarly $q'^*\!\ff'$ and $q^*\!\ff$
are the pull-back sections relative to the 
projections  
$q\,:\,Z^{[3]}\rightarrow Y^{[3]}$ and  $q'\,:\,Z^{[3]}\rightarrow 
Y'^{[3]}$.
\end{definition}
The relation of stable isomorphism is an equivalence relation.

The bundle gerbe $(\EE,Y,M,\ff)$ is called trivial if it is stably
isomorphic to the trivial bundle gerbe $(Y\times H,Y,M,\11)$; we thus have
that $\EE$ and $\NN_1^{-1}\NN_2$ are isomorphic as 
$H$-bibundles, i.e.
\eq
\EE \sim \NN_1^{-1}\NN_2
\en
and that $\ff=\vvphi_1^{-1}(\ll_{13}^{-1}\ll_{23}\ll_{12})$ 
where $\ll$ is the global central section of 
$\eta^{-1}\equiv\NN_2\EE^{-1}\NN_1^{-1}$.

\begin{proposition}\label{Note1}
Consider a bundle gerbe ${\GG}=(\EE,Y,M,\ff)$ with
submersion 
$\wp\,:\,Y\rightarrow M$; a new submersion
$\wp'\,:\,Y'\rightarrow M$ and a (smooth) map
$\sigma\,:\,Y'\rightarrow Y$ 
compatible with $\wp$ and $\wp'$ (i.e. $\wp\circ\sigma=\wp'$). 
The pull-back bundle gerbe $\sigma^*{\GG}$ 
(with obvious abuse of notation) is given by
$(\sigma^*_{\!\!\!\!}\EE,Y',M,\sigma^*\!\ff)$.  We have that the bundle gerbes ${\GG}$
and $\sigma^*{\GG}$ are stably equivalent.
\end{proposition}
\begin{proof}
Consider the following identity on $Y^{[4]}$:
\eq
\EE_{11'}\EE_{1'2'}\EE^{-1}_{22'}=\eta_{12}\EE_{12}
\label{stable}
\en
where $\eta_{12}=\TT_{11'2'}\TT^{-1}_{122'}$ so that $\eta_{12}^{-1}$ has
section $\ll_{12}=\ff^{-1}_{122'}\ff_{11'2'}\,$; the labelling $1,1',2,2'$ 
instead of $1,2,3,4$ is just a convention. 
Multiplying three times
(\ref{stable}) we obtain the following identity between trivial
bundles on $Y^{[6]}$
$\,
\EE_{11'\,}\TT_{1'2'3'\,}\EE_{11'}^{-1}=
\eta_{12}\,\EE_{12}\eta_{23}\EE_{12}^{-1}\,\TT_{123\,}\eta^{-1}_{13}
$.
The sections of (the inverses of) these bundles satisfy
\eq
\vvphi_{11'}(\ff')=
\ll_{13}^{-1}\ff\vvphi_{12}(\ll_{23})\ll_{12}
\label{stable1}~,
\en
thus $\EE_{1'2'}$ and 
$\EE_{12}$ give stably equivalent bundle gerbes.
Next we pull-back the bundles in (\ref{stable})
using  $(id,\sigma,id,\sigma)\,:\,Z^{[2]}
\rightarrow Y^{[4]}$ where $Z=Y\times_M Y'$; 
recalling that the product commutes
with the pull-back we obtain relation (\ref{stably}) with
$\eta=(id,\sigma,id,\sigma)^*\eta_{12}$ and $N=(id,\sigma)^*\EE$.
We also pull-back (\ref{stable1}) with 
$(id,\sigma,id,\sigma,id,\sigma)\,:\,Z^{[3]}\rightarrow
Y^{[6]}$ and obtain formula (\ref{stably1}).
\end{proof}

\begin{theorem}\label{loctriv}
Locally a bundle gerbe is always trivial: 
$\forall x\in M$ there is an open $O$ of $x$ such that
the bundle gerbe restricted to $O$ is stably isomorphic to the trivial
bundle gerbe $(Y_{\,\,}\!|_{O}^{[2]}
\times H,Y_{\,\,}\!|_{O},O,\11)$. 
Here  $Y_{\,\,}\!|_{O}$ is $Y$ restricted to $O_{}\/$:
 $\,Y_{\,\,}\!|_{O}=\{y\in Y\,|\,\wp(y)\in O\subset M\}$.
Moreover in any sufficiently small open $\U$ of $\YO^{[3]}$ one has
\eq
\ff=\sspp_{\!\!\!\!13}\ssp_{23}^{-1}\ss_{12}^{-1}
\label{localtriv}
\en
with $\ss_{12}^{-1}, \ssp_{\!\!23}^{-1}$ and $\sspp_{\!\!\!\!13}$ respectively 
sections of $\EE_{12}^{-1}, \EE_{23}^{-1}$ and $\EE_{13}$ that 
are pull-backs of sections of $\EE$.
\end{theorem}
\begin{proof} 
Choose  $O\subset M$ such that there exists a section 
$\sigma \,:\,O\rightarrow \YO$. Define the maps
\eqa
&&r_{[n]}\,:\,\YO^{[n]}\rightarrow \YO^{[n+1]}\nonumber\\
&&(y_1,\ldots y_n)\mapsto (y_1,\ldots y_n, \sigma(\wp(y_n)))
\nonumber
\ena
notice that
$\sigma(\wp(y_1))=\sigma(\wp(y_2))\ldots=\sigma(\wp(y_n))$.
It is easy to check the following equalities between maps on
$\YO^{[2]}$, 
$
{}~p_{12}\circ r_{[2]}=id~,~~p_{13}\circ r_{[2]}=r_{[1]}\circ p_1~,~~
p_{23}\circ r_{[2]}=r_{[1]}\circ p_2\:,
$
and between maps on $\YO^{[3]}$
\eq
p_{123}\circ r_{[3]}=id~,~~p_{124}\circ r_{[3]}=r_{[2]}\circ p_{12}~,~~
p_{234}\circ r_{[3]}=r_{[2]}\circ p_{23}~,~~
p_{134}\circ r_{[3]}=r_{[2]}\circ p_{13}~.
\label{r3p}
\en
We now pull back with $r_{[2]}$ the identity 
$\EE_{12}=\TT\EE_{13}\EE_{23}^{-1}$
and obtain the following local trivialization of $\EE$
$$
\EE=r_{[2]}^*(\TT)\,{\NN}_{1\,}{\NN}_2^{-1}
$$
where ${\NN}_1=p_1^*(N)$,  ${\NN}_2=p_2^*(N)$ and
$N=r^*_{[1]}(\EE)$.
Let $\U=U\!\times_OU'\!\times_O U'' \subset \YO^{[3]}$ where
$U,U',U''$ are opens of $\YO$ that respectively admit the sections 
${\nn}:\,U\rightarrow N$, ${\nnp}:\,U'\rightarrow N$, 
${\nnpp}:\,U''\rightarrow N$. 
Consider the local sections $\ss=r^*_{[2]}(\ff^{-1})\nn_1\nnp_{\!\!2}^{-1}:\,
U\!\times_O^{\!}U'\rightarrow \EE$,
$\ssp=r^*_{[2]}(\ff^{-1})\nnp_{\!\!2}\nnpp_{\!\!\!\!3}^{-1}\!\!:\,
U'\!\times_O^{\!}U''\rightarrow \EE$,  
$\;\sspp=r^*_{[2]}(\ff^{-1})\nn_1\nnpp_{\!\!\!\!3}^{-1}\!\!:\,
U\!\times_O^{\!}U''\rightarrow \EE$ and pull them back to local sections
$\ss_{12}$ of $\EE_{12}$, $\ssp_{\!\!23}$ of $\EE_{23}$ and
$\sspp_{\!\!\!\!13}$ of $\EE_{13}$. Then (\ref{localtriv}) holds because,
using (\ref{r3p}), the product  
$\sspp_{\!\!\!\!13}\ssp_{\!\!23}^{-1}\ss_{12}^{-1}$ equals the pull-back 
with $r_{[3]}$ of the section 
$\ff^{-1}_{134}\ff_{124}\vvphi_{12}(\ff_{234})=\ff_{123}\,$ 
[cf. (\ref{cocycleglobal})].
\end{proof}

\subsection*{Local description}
\noindent \\
Locally we have the following description of a bundle gerbe;
we choose an atlas of charts for the bundle $\EE$ on  
$Y^{[2]}$, i.e. sections $\tt^i\,:\,\U^i\rightarrow \EE$
relative to a trivializing covering $\{\U^i\}$ of $Y^{[2]}$.
We write $\EE=\{h^{ij},\vphi^i\}$. We choose also atlases 
for the pull-back bundles 
$\EE_{12}, \EE_{23}, \EE_{13}$; we write 
$\EE_{12}=\{h_{12}^{ij},\vphi_{12}^i\}$, 
$\EE_{23}=\{h_{23}^{ij},\vphi_{23}^i\}$, 
$\EE_{13}=\{h_{13}^{ij},\vphi_{13}^i\}$, 
where these atlases are relative to  
a common trivializing covering $\{\U^i\}$ of $Y^{[3]}$. 
It then follows that $\TT=\{f^i{f^j}^{-1},Ad_{f^i}\}$ 
where $\{{f^i}^{-1}\}$ are the local representatives for
the section $\ff^{-1}$ of $\TT$.
We also consider atlases for the bundles on $Y^{[4]}$ that are
relative to a common trivializing covering $\{\U^i\}$ of $Y^{[4]}$
(with abuse of notation we denote with the same index $i$
all these different coverings\footnote{An explicit construction is for
example obtained pulling back the atlas of $\EE$ to the pull-back
bundles on $Y^{[3]}$ and on $Y^{[4]}$. 
The sections $\tt^i\,:\,\U^i\rightarrow \EE$ induce
the associated sections
$\tt^i_{12}\equiv p^*_{12}(\tt^i)\,:\,\U^i_{12}\rightarrow \EE_{12}$
where $p_{12}\,:\,Y^{[3]}\rightarrow Y^{[2]}$ and 
$\U^i_{12}\equiv
p_{12}^{-1}(\U^i)$. We then have $\EE_{12}=\{h_{12}^{ij},\vphi_{12}^i\}$ 
with $h_{12}^{ij}=p^*_{12}(h^{ij})$, $\,\vphi_{12}^i=p^*_{12}(\vphi^i)$.
Similarly for $\EE_{13}$, $\EE_{23}$. 
The $Y^{[3]}$ covering given by 
the  opens 
$\U^I \equiv \U^{ii'i''}\equiv{\U_{12}^i\cap 
\U_{23}^{i'}\cap \U_{13}^{i''}}$ 
can then be used for a common trivialization of the $\EE_{12}$,
$\EE_{13}$ and $\EE_{23}$ bundles; the respective sections
are $\tt^I_{12}=\tt^i_{12}|_{_{\U^I}}$, 
$\tt^I_{23}=\tt^{i'}_{23}|_{_{\U^I}}$, 
$\tt^I_{13}=\tt^{i''}_{13}|_{_{\U^I}}$; similarly for the 
transition functions
$h_{12}^I, h_{23}^I, h_{13}^I$
and for $\vphi^I_{12},\vphi^I_{23},\vphi^I_{13}$.
In ${\U^I}$ we then have $\ff^{-1}={f^I}^{-1\,}\tt^I_{12}\tt^I_{23\,}
{\tt^I_{13}}^{\!\!\!-1}\,.$
}).  
Then  (\ref{trivialT}), that we rewrite as $\EE_{12}\EE_{23}=
\TT\EE_{13}$, reads
\eq
h^{ij}_{12}\vphi^j_{12}(h^{ij}_{23})=f^ih^{ij}_{13}{f^j}^{-1}~~,~~~
\vphi^i_{12}\circ \vphi^i_{23}=Ad_{f^i}\circ \vphi^i_{13} \label{localbg1}
\en
and relation (\ref{cocycleglobal}) reads
\eq
\vphi^i_{12}(f^i_{234})f^i_{124}=f^i_{123}f^i_{134} \label{localbg2}~.
\en

\subsection*{Bundles and local data on \mbox{{\boldmath $M$}}}
\noindent \\
Up to equivalence under stable isomorphisms, there is an alternative
geometric description of bundle gerbes, in terms of 
bundles on $M$. 
Consider the sections $\sigma_\al\;:\;O_\al\rightarrow Y$,
relative to a covering $\{O_\al\}$ of $M$ and consider also 
the induced sections
$(\sigma_\al, \sigma_\be)
\;:\;O_{\al\be}\rightarrow Y^{[2]}$,
$(\sigma_\al,\sigma_\be,\sigma_\ga)\;:\;O_{\al\be\ga}\rightarrow
Y^{[3]}$.
Denote by 
$\EE_{\al\be}$, $\TT_{\al\be\ga}$ the pull-back of the $H$-bibundles 
$\EE$ and $\TT$ via $(\sigma_\al,\sigma_\be)$ and
$(\sigma_{\al}, \sigma_{\be},\sigma_{\ga})$. Denote also by
$\ff_{\al\be\ga}$ the pull-back of the section $\ff$.
Then, following Hitchin description of abelian gerbes,

\begin{definition} A gerbe is a 
collection $\{\EE_{\al\be}\}$ of $H$ principal bibundles $\EE_{\al\be}$ on each
$O_{\al\be}$ such that on the triple intersections $O_{\al\be\ga}$ 
the product bundles $\EE_{\al\be}\EE_{\be\ga}\EE_{\al\ga}^{-1}$ are 
trivial, and such that on the quadruple intersections
$O_{\al\be\ga\delta}$  we have 
$\ff_{\al\be\delta}\vvphi_{\al\be}(\ff_{\be\ga\delta})
=\ff_{\al\ga\delta}\ff_{\al\be\ga}\,.$ \label{Hi}
\end{definition}
We also define two gerbes, given respectively by
$\{\EE'_{\al\be}\}$ and $\{\EE_{\al\be}\}$ (we can always consider a
common covering $\{O_\al\}$ of $M$),  
to be stably equivalent if there exist bibundles ${\NN}_{\al}$
and trivial bibundles $\eta_{\al\be}$ with (global central) 
sections $\ll^{-1}_{\al\be}$ such that
\eqa
& &{\NN}_{\al}\EE'_{\al\be}{\NN}_{\be}^{-1}=\eta_{\al\be}\EE_{\al\be}~,
\label{stale}\\
& &\vvphi_{\al}(\ff'_{\al\be\ga})=
\ll_{\al\ga}^{-1}\ff_{\al\be\ga}\vvphi_{\al\be}(\ll_{\be\ga})\ll_{\al\be}
\label{stale1}~.
\ena

A local description of the $\EE_{\al\be}$ 
bundles in terms of the local data 
(\ref{localbg1}), (\ref{localbg2}) can be given considering
the refinement $\{O^i_\al\}$ of the $\{O_\al\}$ cover of $M$
such that $(\sigma_\al,\sigma_\be)(O^{ij}_{\al\be})
\subset \U^{ij}\subset Y^{[2]}$, the refinement  $\{O^i_\al\}$ such that
$(\sigma_\al,\sigma_\be,\sigma_\ga)(O^{ijk}_{\al\be\ga})
\subset\U^{ijk}\subset Y^{[3]}$, and similarly for $Y^{[4]}$. 
We can then define the local data on $M$ 
\eqa
& h^{ij}_{\al\be}\;:\;O^{ij}_{\al\be}\rightarrow H ~~~~~~~~~~~~~~~~ 
& \vphi^i_{\al\be}\;:\;O^i_{\al\be}\rightarrow Aut(H) \nonumber \\[.5 em]  
&~~~h^{ij}_{\al\be}=h^{ij}_{12}\circ(\sigma_\al,\sigma_\be) ~~~~~~~~~~~~~
&\vphi^i_{\al\be}=\vphi_{12}^i\circ (\sigma_\al,\sigma_\be) 
\ena
and
\eqa 
& &f^i_{\al\be\ga}\;:\;O^i_{\al\be\ga}\rightarrow H\nonumber\\[.5 em]
& &f^i_{\al\be\ga}=f^i\circ(\sigma_{\al},\sigma_{\be},\sigma_{\ga})~.
\ena
It follows that
$\EE_{\al\be}=\{h^{ij}_{\al\be},\vphi^i_{\al\be}\}$
and 
$\TT_{\al\be\ga}=\{f^i_{\al\be\ga}{f_{\al\be\ga}^{j\,\,-1}}, 
Ad_{f^i_{\al\be\ga}}\}$.
Moreover relations  (\ref{localbg1}), (\ref{localbg2}) imply the
relations between local data on $M$
\eq
h^{ij}_{\al\be}\vphi^j_{\al\be}(h^{ij}_{\be\ga})
=f^i_{\al\be\ga}h^{ij}_{\al\ga}
{f^j_{\al\be\ga}}^{\!\!\!\!\!\!-1}\,,\label{localg11}
\en
\eq
\vphi^i_{\al\be}\circ \vphi^i_{\be\ga}=Ad_{f^i_{\al\be\ga}}\circ 
\vphi^i_{\al\ga} \,,  \hskip 0.5cm
\vphi^i_{\al\be}(f^i_{\be\ga\delta})f^i_{\al\be\delta}=f^i_{\al\be\ga}
f^i_{\al\ga\delta} \label{localg22}\,.~
\en
We say that (\ref{localg22}) define a nonabelian 
\v{C}ech $2$-cocycle. 
{}From (\ref{stale}), (\ref{stale1}) we see that 
two sets $\{h^{ij}_{\al\be},\vphi_{\al\be}^i,f^i_{\al\be\ga}\}$,
$\{h'^{ij}_{\al\be},\vphi'^i_{\al\be},f'^i_{\al\be\ga}\}$
of local data on $M$ are stably isomorphic if
\begin{eqnarray}
&&h^{ij}_{\al\,}\vphi^j_{\al}(h'^{ij}_{\al\be})_{\,}
\vphi_\al^{j\,}\vphi'^j_{\al\be\,}{\vphi_\be^j}^{\!-1}(h^{ij}_\be)=
\ell_{\al\be\,}^ih^{ij}_{\al\be\,}{\ell^j_{\al\be}}^{\!\!-1}
\,,\\
&&\vphi_\al^i\circ\vphi'^i_{\al\be}\circ {\vphi^i_{\be}}^{\!-1}
=Ad_{\ell^i_{\al\be}}\circ \vphi^i_{\al\be} 
\label{localg1}\,,\\
&&\vphi_{\al}^i(f'^i_{\al\be\ga})=\ell^i_{\al\be\,}\vphi^i_{\al\be}
(\ell^i_{\be\ga})f^i_{\al\be\ga}
\ell^{i\,\,-1}_{\al\ga} \label{staseonM}\,,
\end{eqnarray}
here
${\NN}_\al=\{h^{ij}_\al,\vphi_{\al}\}\,,~
\EE'_{\al\be}=\{h'^{ij}_{\al\be},\vphi'_{\al\be}\}$ and
$\eta_{\al\be}=\{\ell^{i}_{\al\be}{\ell^{j}_{\al\be}}^{\!\!\!\!-1},
Ad_{\ell^{i}_{\al\be}}\}$.
\sk
\sk
We now compare the gerbe
$\{\EE_{\al\be}\}$ 
obtained from a bundle gerbe $\GG$ using the sections
$\sigma_\al\,:\,O_\al\rightarrow Y$ 
to the gerbe $\{\EE'_{\al\be}\}$ 
obtained from $\GG$ using a different choice of
sections $\sigma'_\al\,:
\,O_\al\rightarrow Y$. 
We first pull back the bundles in (\ref{stable})
using  $(\sigma_\al,\sigma'_\al,\sigma_{\be},\sigma'_\be)\,:\,
O_{\al\be}\rightarrow Y^{[4]}$; recalling that the product commutes
with the pull-back we obtain the following relation between bundles
respectively on 
$O_{\al},\,O_{\al\be},\,O_{\be}$ and on $O_{\al\be},\,O_{\al\be}$\,,
\[
{\NN}_{\al} \EE'_{\al\be}{\NN}_{\be}^{-1}=\eta_{\al\be}\EE_{\al\be}~,
\]
here ${\NN}_{\al}$  equals the pull-back of $\EE_{11'}$ 
with  $(\sigma_\al,\sigma'_{\al})\,:\,O_{\al}\rightarrow Y^{[2]}$.
We then pull back (\ref{stable1}) with 
$(\sigma_\al,\sigma'_\al,\sigma_\be,
\sigma'_{\be},\sigma_\ga,\sigma'_\ga)\,:\,O_{\al\be\ga}\rightarrow
Y^{[6]}$ and obtain formula (\ref{stale1}). Thus $\{\EE'_{\al\be}\}$ and
$\{\EE_{\al\be}\}$  are stably equivalent gerbes.
We have therefore shown that  the equivalence class of a gerbe 
(defined as a collection  of bundles on $O_{\al\be}\subset M$)
is independent from the choice of sections 
$\sigma_\al\,:\,O_\al\rightarrow Y$ used to obtain 
it as pull-back from a bundle gerbe.
\sk
It is now easy to prove that equivalence classes of bundle
gerbes are in one to one correspondence with equivalence classes of
gerbes $\{\EE_{\al\be}\}$, and therefore with equivalence classes of 
local data on $M$.
First of all we observe that a bundle gerbe $\GG$ and its pull-back
$\sigma^*{\GG}=(\sigma^*_{\!\!\!\!}\EE,Y',M,\sigma^*\!\ff)$ (cf. Theorem
$\ref{Note1}$)
give the same gerbe $\{\EE_{\al\be}\}$ if we use
the sections $\sigma'_\al\,:\,O_{\al}\rightarrow Y'$ for 
$\sigma^*{\GG}$ and the sections $\sigma\circ\sigma'_\al
\,:\,O_{\al}\rightarrow Y$ for ${\GG}$. 
It then follows that two stably equivalent bundle gerbes give two 
stably equivalent gerbes. In order to prove the converse we associate
to each gerbe $\{\EE_{\al\be}\}$ a bundle gerbe and then we prove that
on equivalence classes this operation is the
inverse of the operation 
$\GG\rightarrow \{\EE_{\al\be}\}$.
Given 
$\{\EE_{\al\be}\}$ we consider $Y=\sqcup O_\al$, the disjoint
union of the opens $O_{\al}\subset M$, with projection $\wp(x,\al)=x$.
Then $Y^{[2]}$ is the disjoint union of the opens $O_{\al\be}$, 
i.e. $Y^{[2]}=\sqcup O_{\al\be}=\cup O_{\al,\be}$, where   
$O_{\al,\be}=\{(\al,\be,x)\, / \,x\in O_{\al\be}\}$, similarly  
$Y^{[3]}=\sqcup O_{\al\be\ga}=\cup O_{\al,\be,\ga}$ etc..
We define $\EE$ such that $\EE|_{{O_{\al,\be}}}=\EE_{\al\be}$ 
and we define the section $\ff^{-1}$ of
$\TT=\EE_{12}\EE_{23}\EE^{-1}_{13}$ 
to be given by
$\ff^{-1}|_{_{O_{\al,\be,\ga}}}=\ff^{-1}_{\al\be\ga}$, 
thus
(\ref{cocycleglobal}) holds. We write
$(\sqcup\EE_{\al\be},\sqcup O_\al,M, 
\sqcup \ff_{\al\be\ga})$ for this bundle gerbe. 
If we pull it back with
$\sigma_\al\,:\,O_\al\rightarrow Y$, $\sigma_\al(x)=(x,\al)$ we obtain
the initial gerbe $\{\EE_{\al\be}\}$. 
In order to conclude the proof we have to show that
$(\sqcup \EE_{\al\be},\sqcup O_\al,M,
\sqcup \ff_{\al\be\ga})$ 
is stably isomorphic to the bundle gerbe
${\GG}=(\EE,Y,M,\ff)$ if $\{\EE_{\al\be}\}$ is
obtained from   ${\GG}=(\EE,Y,M,\ff)$ 
and sections $\sigma_\al\,:\,O_\al\rightarrow Y$.
This holds because 
$(\sqcup \EE_{\al\be},\sqcup O_\al,M,
\sqcup \ff_{\al\be\ga})=\sigma^*{\GG}$
with $\sigma\,:\,\sqcup O_\al\rightarrow Y$ given by
$\sigma|_{_{O_\al}}=\sigma_\al$.

\sk

{}We end this section with a comment on normalization.  
There is no loss in generality if we
consider  for all $\al,\be$  and for all $i$  
\eq
\vphi_{\al\al}^i=id~~,~~~f^i_{\al\al\be}=1~~,~~~f^i_{\al\be\be}=1
\en
Indeed first notice from  (\ref{localg11}) and (\ref{localg22})
that $\vphi_{\al\al}^i=Ad_{f^i_{\al\al\al}}$ and 
$\vphi^i_{\al\al}(f^i_{\al\al\be})=f^i_{\al\al\al}$ so that
$f^i_{\al\al\be}=f^i_{\al\al\al}|_{_{O_{\al\be}}}$.
Now, if $f^i_{\al\al\al}\not=1$ consider the stably equivalent
local data obtained from
$\EE'_{\al\be}\equiv\eta_{\al\be}\EE_{\al\be}$
where $\eta_{\al\be}=\{\ell^{i}_{\al\be}{\ell^{j}_{\al\be}}^{\!\!\!\!-1},
Ad_{\ell^{i}_{\al\be}}\}$ with 
$\ell^i_{\al\be}={f^i_{\al\al\al}}^{\!\!\!\!\!\!\!-1}|_{_{O_{\al\be}}}$. 
{}From (\ref{localg1}) we have 
$\vphi'^i_{\al\al}=id$;
{}from (\ref{staseonM}) we have $f'^i_{\al\al\be}=1$, it then also follows 
$f'^i_{\al\be\be}=1$. 
\sk\sk


%

\sk
\section{\bf Nonabelian Gerbes from Groups Extensions }

We here associate a bundle gerbe on the manifold  $M$ 
to every group extension
\eq
1\rightarrow H\rightarrow E\stackrel{{\pi}}{{\rightarrow}} G\rightarrow 1 \label
{ext}
\en
and left principal $G$ bundle $P$ over $M$. We identify $G$ with 
the coset ${}_{\mbox{\small $\small H$}}\!\!\setminus\!E$
so that $E$ is a left $H$ principal bundle. $E$ is naturally a bibundle,
the  right action too is given by the group law in $E$
\eq
e\triangleleft h=eh=(ehe^{-1})_{} e \label{109}
\en
thus  $\vphi_e(h)=ehe^{-1}$. 
We denote by $\tau\,:\,P^{[2]}\rightarrow G$,  $~\tau(p_1,p_2)=g_{12}$ 
the map that associates to any two points $p_1,p_2$ of $P$ that 
live on the same fiber the unique element $g_{12}\in G$ such that 
$p_1=g_{12} p_2$.
Let  $\EE\equiv\tau^*(E)$ be the pull-back of $E$ on $P^{[2]}$, explicitly
$\EE=\{(p_1,p_2;e)~|~\pi(e)=\tau(p_1,p_2)=g_{12}\}$. Similarly
$\EE_{12}=\{(p_1,p_2,p_3;e)~|~\pi(e)=\tau(p_1,p_2)=g_{12}\}$, for brevity of 
notations we set $e_{12}\equiv (p_1,p_2,p_3;e)$. Similarly with $\EE_{23}$ and
$\EE_{13}$, while $e_{13}^{\,-1}$ is a symbolic notation for a 
point of $\EE_{13}^{-1}$. Recalling (\ref{action-1}) we have 
\eq
{}~~~~~(he)_{13}^{\,-1}=(ek)_{13}^{\,-1}=
k^{-1}\,e_{13}^{\,-1}~~~~,~~~~~e_{13}^{\,-1}\triangleleft^{\!\!-1}h
=k\,e_{13}^{\,-1}\label{50}
\en
where $k=e^{-1}he\,$. We now consider the point  
\eq
\ff^{-1}(p_1,p_2,p_3)
\equiv[e_{12},e'_{23},(ee')_{13}^{-1}]\,\in \,\EE_{12}\EE_{23}\EE^{-1}_{13}
\label{thesec}
\en  
where the square bracket denotes, as in (\ref{13}), the equivalence class 
under the $H$ action\footnote{It can be shown that a realization of the
equivalence class $[e_{12},e'_{23}]\in \EE_{12}\EE_{23}$ is given by 
$(p_1,p_2,p_3;ee')$ where $ee'$ is just the product in $E$. (We won't
use this property).}.
Expression (\ref{thesec}) is well defined because
$\pi(ee')=\pi(e)\pi(e')=g_{12}g_{23}=g_{13}$ the last equality following from
$p_1=g_{12}p_2\,,~p_2=g_{23}p_3\,,~ p_1=g_{13}p_3\,$. 
Moreover $\ff(p_1,p_2,p_3)$ is independent from $e$ and $e'$,
indeed let $\hat e$ and $\hat e'$ be two other elements of $E$ such that
$\pi(\hat e)=\pi(e)\,,\:\pi(\hat e')=\pi(e')$; then $\hat e=h e$,
$\hat e'=h' e'$ with $h,h'\in H$ and 
$[{\hat e}_{12},{\hat e'}_{23},(\hat{e}\hat e')_{13}^{-1}]=
[h\,{e}_{12},h'\,{e'}_{23},e'^{-1}h'^{-1}e^{-1}h^{-1}ee'\,(ee')_{13}^{-1}]=
[e_{12},e'_{23},(ee')_{13}^{-1}]$. This shows that (\ref{thesec}) defines
a global section $\ff^{-1}$ of $\TT\equiv \EE_{12}\EE_{23}\EE^{-1}_{13}$.
Using the second relation in (\ref{50}) we also have that $\ff^{-1}$ is 
central so that $\TT$ is a trivial bibundle. Finally (the inverse of) 
condition 
(\ref{cocycleglobal}) is easily seen to hold and we conclude that
$(\EE,P,M,\ff)$ is a bundle gerbe. It is the so-called {\it lifting bundle
gerbe}.
\sk
\sk

\sk
\section{\bf Bundle Gerbes Modules }
The definition of a module for a nonabelian bundle gerbe is inspired by
the abelian case \cite{Bouwknegt:2001vu}.

\begin{definition}\label{module}
Given an H-bundle gerbe $(\EE, Y, M, \ff)$, an $\EE$-module consists of a
triple
$(\QQ,\ZZ,\zz)$ where $\QQ\rightarrow Y$ is a $D$-$H$ bundle, $\ZZ \rightarrow Y^{[2]}$ is 
a trivial $D$-bibundle and $\zz$ is a global central section of $\ZZ^{-1}$
such that:
\newline
i) on $Y^{[2]}$  
\eq
\QQ_1\EE = \ZZ\QQ_2 \label{mod0}
\en 
and moreover 
\eq
\vvphi_{12} = \vvpsi_1^{-1}\circ\,^{\bar{z}^{-1}_{12}}\circ \vvpsi_2. \label{mod1}
\en
\newline
ii) (\ref{mod0}) is compatible with the bundle gerbe structure of 
$\EE$, i.e. from (\ref{mod0}) we have  
$\QQ_1\TT=\ZZ_{12}\ZZ_{23}\ZZ_{13}^{-1}\QQ_1$ on $Y^{[3]}$ 
and we require that
\eq
\zz_{23}\zz_{12} = \zz_{13} \vvxi_1(\ff) \label{mod}
\en
holds true.
\end{definition}
\noindent
\begin{remark}
Let us note that the pair $(\ZZ,\zz^{-1})$ and the pair $(\TT,\ff^{-1})$ 
in the above definition give the isomorphisms 
\eq
z:\QQ_1\EE \rightarrow \QQ_2 ~~~\:\:,\:\:~~~~~~f:\EE_{12}\EE_{23}\rightarrow \EE_{13}
\en
respectively of $D$-$H$ bundles on $Y^{[2]}$ and of bibundles on $Y^{[3]}$.
Condition {\it ii)} in Definition $16$ is then equivalent to the 
commutativity of the following diagram  
\begin{center}
\eq
\begin{CD}
\QQ_1 \EE_{12}  \EE_{23}@>{id}_1 {f}>>
        \QQ_1 \EE_{13} \\
@V{z}_{12\,}{id}_3 VV   @VV {z}_{13}V \\ \label{diagram1}
\QQ_2 \EE_{23} @>{z}_{23}>> \QQ_3
\end{CD}
\en
\end{center}
\end{remark}

\begin{definition}
We call two bundle gerbe modules $(\QQ,\ZZ,\zz)$ and
$(\QQ',\ZZ',\zz')$ (with the same crossed module structure) 
equivalent if: 
\newline
i) $\QQ$ and $\QQ'$ are isomorphic as  $D$-$H$
bundles; we write $\QQ=\II\QQ'$ where the $D$-bibundle $\II$ 
has global central section $\ii^{-1}$ and $\vvpsi = \,^{\bar{i}^{-1}}\circ \vvpsi'$ 
\newline
ii) the global central
sections $\zz$, $\zz'$ and $\ii^{-1}$ satisfy the condition
$\zz'_{12}= \ii_2^{-1}\zz_{12}\ii_1$.
\end{definition}
Let us now assume that 
we have two stably equivalent bundle gerbes $(\EE,Y,M,\ff)$
and $(\EE',Y',M,\ff')$ 
with $Y'=Y$. We have [cf. (\ref{stably}), (\ref{stably1})]
$\eta_{12}\EE_{12} = \NN_1\EE^{'}_{12}\NN_2^{-1}$ and $
\vvphi_1(\ff')=
\ll_{13}^{-1}\ff\vvphi_{12}(\ll_{23})\ll_{12}$.
Let $\QQ$ be an $\EE$-module and $\II$ a trivial $D$-bibundle with a 
global central
section $\ii^{-1}$. It is trivial  to 
check that $\II\QQ\NN$ is an $\EE'$-module with $\ZZ'_{12} =
\II_1\vvxi_1(\eta_{12})\ZZ_{12}\II^{-1}_2$ and $\zz'_{12} = \ii_2^{-1}
\zz_{12}\vvxi(\eta_{12})\ii_1$. It is now easy to compare modules of stably
equivalent gerbes that in general have $Y\not=Y'$.
\begin{proposition}
Stably equivalent gerbes have the same equivalence classes of modules. 
\end{proposition}

Now we give the description of bundle gerbes modules in terms of local data on $M$.
Let $\{EE_{\alpha \beta}\}$ be a gerbe in the sense of definition \ref{Hi}.
\begin{definition}
A module for the gerbe $\{\EE_{\alpha \beta}\}$ is given by a collection
$\{\QQ_{\alpha}\}$
of $D$-$H$ bundles such that on
double intersections $O_{\alpha \beta}$ there exist trivial $D$-bibundles
$\ZZ_{\alpha \beta}$,
$\QQ_{\alpha}\EE_{\alpha \beta} = \ZZ_{\alpha \beta}\QQ_{\beta}$,
with global central 
sections $\zz_{\alpha \beta}$ of $\ZZ^{-1}_{\alpha \beta}$ such that
on triple intersections $O_{\alpha \beta \gamma}$
\eq
\zz_{\beta \gamma}\zz_{\alpha \beta} = \zz_{\alpha \gamma} \vvxi_\alpha(\ff_{\alpha
\beta \gamma}) \label{locmod}
\en
and on double intersections $O_{\alpha \beta}$
\eq
\vvphi_{\alpha \beta} = \vvpsi_{\alpha}^{-1}\circ\,^{\bar{z}^{-1}_{\alpha \beta}}\circ
\vvpsi_{\beta}. \label{locmod1}
\en
\end{definition}

\subsection*{Canonical module}
\noindent \\
For each $H$-bundle gerbe $(\EE,Y,M,\ff)$ we have a canonical
$\EE$-module associated with it; it is constructed as follows.
As a left $Aut(H)$-bundle
the canonical module is simply the trivial bundle over 
$Y$. The right action of $H$ is given by the canonical homomorphism  $Ad: H\rightarrow
Aut(H)$. For $(y, \eta) \in Y\times Aut(H)$ we have $\vvxi_{(y,\eta)}(h) =
\eta\circ Ad_h \circ \eta^{-1}=Ad_{\eta(h)}$and $\vvpsi_{(y,\eta)}(h) = \eta(h)$. 
The $Aut(H)$-$H$ bundle morphism $z:(Y\times Aut(H))_1\EE \rightarrow
(Y\times Aut(H))_2$ is given in the following way. A generic
element of $(Y\times Aut(H))_1\EE$ is of the form
$[(y,y',(y,\eta)),e]$ 
where $\eta \in Aut(H), (y,y') \in Y^{[2]}$and $e\in \EE$ such that
$p_1\circ p(e)=y$ and $p_2\circ p(e)=y'$. Here $p$ is the projection
$p: \EE \rightarrow Y^{[2]}$.
We set
$$
z([(y,y',(y,\eta)),e])=(y, y',(y',\eta\circ\vvphi_e)).
$$
The commutativity of diagram (\ref{diagram1}) is
equivalent to the following statement 
$$
\eta\circ\vvphi_{f[e_1,
  e_2]}=\eta\circ\vvphi_{e_1}\circ\vvphi_{e_2} 
$$
and this is a consequence of the isomorphism of $H$-bibundles
$$
f:\EE_{12}\EE_{23}\rightarrow\EE_{13}.
$$
We have
$$
f([e_1,e_2]h)=(f[e_1,e_2])h\,,
$$
but we also have
$$
f([e_1, e_2)]h)=f(\vvphi_{e_1}\circ
\vvphi_{e_2}(h)[e_1,e_2])=\vvphi_{e_1}\circ
\vvphi_{e_2}(h)f([e_1, e_2]).
$$
On the other hand we can write
$$
(f[e_1,e_2])h=\vvphi_{f[e_1,e_2]}(h)f[e_1, e_2].
$$
Hence
$$
\vvphi_{f[e_1, e_2]}(h)=\vvphi_{e_1}\circ
\vvphi_{e_2}(h)
$$
and the commutativity of diagram (\ref{diagram1}) follows. 
We denote the canonical module as $can$ in the following. 

\sk

In the case of a bundle gerbe $\EE$ associated with the lifting of a 
$G$-principal bundle $P$, as
described in Section $5$, we have another natural module.
We follow the notation of Section $5$. In the exact sequence of groups
(\ref{ext})
$$
1\rightarrow H\rightarrow E\stackrel{{\pi}}{{\rightarrow}} G\rightarrow
1\,,
$$
$H$ is a normal subgroup. This gives the group $H$ the structure of a crossed
$E$-module.

The $\EE$-module $\QQ$ is simply the trivial $E$-$H$ bundle $P\times E \rightarrow P$. 
The $D$-$H$ bundle morphims  $z':\QQ_1\EE\rightarrow \QQ_2$ is given by (recall
$p_1=\pi(\tilde e)p_2$)
$$
z'[(p_1,p_2,(p_1,e),(p_1,p_2,\tilde e)]=(p_1,p_2,(p_2,e\tilde e)),
$$
which of course is compatible with the 
bundle gerbe structure of $\EE$. 
Due to the exact sequence (\ref{ext}) we do have a homomorphism 
$E\rightarrow Aut(\rm{H})$ and hence we have a map 
$$
t: Y\times E\rightarrow Y\times Aut(H),
$$
which is a morphism between the modules compatible with the module structures,
i.e. the following diagram is commutative
\begin{center}
\eq\label{quadrt}
\begin{CD}
\QQ_1\EE_{12} @>z_{12}>>
        \QQ_2 \\
@VtVV   @VVtV \\
can_1\EE_{12} @>z'_{12}>> can_2
\end{CD}
\en
\end{center}  
\sk
\noi More generally given any bundle gerbe $\EE$ and an $\EE$-module $\QQ$ we have 
the trivial $Aut(H)$-$H$ bundle $ Aut(H)\times_{D}\QQ$ (see Section $2$). This gives
a morphism  $t\,:\;\QQ \mapsto can$.

Now suppose that the bundle gerbe $\EE$ is trivialized (stably equivalent to a
trivial bundle gerbe) by $\EE_{12} \sim
\NN^{-1}_1\NN_2$ with $\NN$ an $H$-bibundle on Y, hence a $\EE$-module satisfies
$$
\QQ_2\sim \QQ_1\EE_{12}\sim \QQ_1 \NN^{-1}_1 \NN_2,
$$
hence 
\eq
\QQ_2\NN^{-1}_2 \sim  \QQ_1\NN^{-1}_1 \label{descent}
\en
It easily follows from (\ref{descent}) that $\QQ \NN^{-1}\rightarrow Y$ 
gives descent data for a $D\mbox- H\,\mbox{bundle}$
$\widetilde{Q}$ over $M$. Conversely given a $D$-$H$ bundle
$\widetilde{Q}\rightarrow M$ the bundle $p^*(\widetilde{Q})\NN$
is a $\EE$-module.  This proves the following
\begin{proposition}
For a trivial bundle gerbe $(\EE,Y,M,\ff)$ the category of $\EE$-modules
is equivalent to the category of $D$-$H$ bundles over the base
space $M$. 
\end{proposition}
\sk
\sk
\sk
\section{\bf Bundle Gerbe Connections}

\begin{definition}
A bundle gerbe connection on a bundle gerbe $(\EE,Y,M,\ff)$ is a 
$2$-connection $(\ax, A)$ on $\EE \rightarrow Y^{[2]}$ such that 
\eq\label{sum-cond1}
\ax_{12} \+ \ax_{23}=f^*\ax_{13}\,,
\en
or which is the same
\eq
\ax_{12} \+ \ax_{23} \+ \ax_{13}^r =
\overline{f}^{-1}d\overline{f}+ T_{A_1}(^{}\overline{f}^{-1})
\en 
holds true.
\end{definition}
\sk
In the last equation $\overline{f}^{-1}$ is the bi-equivariant map
$\overline{f}^{-1}: \TT\rightarrow H$ associated with 
the global central section $\ff^{-1}$ of $\TT$. Moreover we used that 
$(\ax^r, A^r)$ is a right $2$-connection on $\EE$ and a left $2$-connection
on $\EE^{-1}$ [cf. (\ref{action-1})].   
\begin{remark}
It follows from (\ref{sum-cond1}) that for a bundle gerbe connection
$A_{12}=A_{13}$ must be satisfied, hence $A$ is a pull-back via $p_1$
on $Y^{[2]}$ of a one form defined on $Y$. We can set $A_1\equiv A_{12}=A_{13}$. 
Definition $21$ contains implicitly the
requirement that $(\ax_{12},\ax_{23})$ are summable, which means that 
$A_1^r = A_2$.
More explicitly (see (\ref{Ar})): 
\eq\label{sum-con2}
{\mathcal A}_{1}+ad_{\axs_{_{^{12}}}}=\vphi_{12}{\mathcal A}_{2}\vphi^{-1}_{12} +
\vphi_{12}d\vphi^{-1}_{12}. 
\en
\end{remark}

The affine sum of bundle gerbe connections is again a bundle gerbe
connection. This is a consequence of the following affine property for
sums of $2$-connection. If on the bibundles $E_1$ and $E_2$ we have two 
couples of summable connections $(\ax_1, \ax_2)$, $(\ax'_1, \ax'_2)$,
then $\lambda\ax_1 +
(1-\lambda)\ax'_1$ is summable to $\lambda\ax_2+(1-\lambda)\ax'_2$ and
the sum is given by
\eq
(\lambda\ax_1 +(1-\lambda)\ax'_1) \+ (\lambda\ax_2+(1-\lambda)\ax'_2)
= \lambda(\ax_1 \+ \ax_2) + (1-\lambda)(\ax'_1 \+ \ax'_2)~. 
\en
We have the following theorem:
\begin{theorem}\label{connectionexists}
There exists a bundle gerbe
connection $(\ax, A)$ on each bundle gerbe $(\EE,Y,M,\ff)$. 
\end{theorem}
\begin{proof}
Let us assume for
the moment the bundle gerbe to be trivial, $\EE
= \NN_1^{-1}\ZZ \NN_2$ with a bibundle $\NN \rightarrow Y$ and a trivial
bibundle $\ZZ \rightarrow Y^{[2]}$ with global central section
$\zz^{-1}$. 
Consider on $\ZZ$ the $2$-connection $(\aalpha,\tilde A)$,
where the $\mbox{Lie}(Aut(H))$-valued one-form $\tilde A$ on $Y$
is the pull-back of a one-form on $M$. Here $\aalpha$ is
canonically determined by $\tilde A$ and $\zz^{-1}$, we have
$\aalpha = \bar{z}^{-1}d\bar{z} + T_A(\bar{z}^{{-1}})$.
Next consider on $\NN$ an arbitrary
$2$-connection $(\tilde{\ax},\tilde A)$.
Since  $\tilde A$ is the pull-back
of a one form on $M$ we have that the sum
$\ax=\tilde{\ax}^r_1 \+ \aalpha \+ \tilde{\ax}_2$ is well defined 
and that $(\ax,A\equiv\tilde A^r)$ is a $2$-connection on $\EE$.
Notice that under the canonical identification 
$\ZZ_{12}\NN_2\NN_2^{-1}\ZZ_{23}=\ZZ_{12}\ZZ_{23}$
we have the canonical identification 
$\aalpha_{12} \+ \tilde{\ax}_2 \+
\tilde{\ax}^r_2 \+ \aalpha_{23}=
 \aalpha_{12} \+ \aalpha_{23}$. The point here is that $\NN_2\NN_2^{-1}$
has the canonical section $\11=[n,n^{-1}]$, 
$n\in \NN_2$, and that  $\tilde{\ax}_2 \+\tilde{\ax}^r_2=
\bar{\11}d\bar{\11}^{-1} + T_A(\bar{\11})$ independently from
$\tilde{\ax}_2$. Then from $\EE = \NN_1^{-1}\ZZ \NN_2$ we have 
$\EE_{12}\EE_{23}\EE_{13}^{-1}=
\NN_1^{-1}\ZZ_{12}\ZZ_{23}\ZZ_{13}^{-1}\NN_1$
and for the connections we have 
\eq
\ax_{12}\+\ax_{23}\+\ax_{13}^r=
\tilde{\ax}^r_1 \+ \aalpha_{12} \+
\aalpha_{23} \+\aalpha_{13}^{r} \+ \tilde{\ax}_1~.
\en
We want to prove that the r.h.s. of this equation equals the
canonical $2$-connection $\overline{f}^{-1}d\overline{f}
+T_{A_1}(^{}\overline{f}^{-1})$ associated with the trivial bundle $\TT$
with section $\ff^{-1}$. We first observe that a similar property 
holds for the sections of $\ZZ_{12}\ZZ_{23}\ZZ_{13}^{-1}$ and
of $\TT$: $\ff^{-1}=
\vvphi^{-1}_1(\zz^{-1}_{12}\zz^{-1}_{23}\zz_{13})
\equiv\nn^{-1}_1 \zz^{-1}_{12}\zz^{-1}_{23}\zz_{13}{\nn_1}$ independently 
from the local section $\nn_1$ of $\NN_1$.  
Then one can explicitly check that this relation implies the relation
$\tilde{\ax}^r_1 \+ \aalpha_{12} \+
\aalpha_{23} \+\aalpha_{13}^{r} 
\+ \tilde{\ax}_1=\overline{f}^{-1}d\overline{f}
+T_{A_1}(^{}\overline{f}^{-1})$.
This proves the validity of the
theorem in the case of a trivial bundle gerbe. 
According to Theorem \ref{loctriv} any gerbe is locally trivial, so
we can use the
affine property of bundle gerbe connections and a partition of unity
subordinate to the covering $\{O_\al\}$ of $M$ 
to extend the proof to arbitrary bundle gerbes.  
\end{proof}

\sk
A natural question arises: can we construct a connection on the bundle gerbe
$(\EE, Y, M, \ff )$ starting with:
\begin{itemize}
\item
its nonabelian \v Cech cocycle $\overline{f}^{-1} : \TT \rightarrow H, \,\, 
\vvphi  : \EE \times H \rightarrow H$
\item
sections $\sigma_{\alpha}:O_{\alpha}\rightarrow Y$ 
\item
a partition of unity $\{\rho_{\alpha}\}$ subordinate to the 
covering $\{O_{\alpha}\}$ of $M$\,?
\end{itemize}
The answer is positive. Let us describe the construction. First we use the local
sections $\sigma_\alpha$ to map $Y_{\,\,}\!|_{O_{\alpha}}^{[2]}$ to $Y^{[3]}$ via
the map $r^{[2]}_{\alpha}:[y,y']\mapsto [\sigma_{\alpha}(x),y, y']$, where
$\wp(y)=\wp(y')=x$, similarly $r^{[1]}_{\alpha}:
Y_{\,\,}\!|_{O_{\alpha}}^{[1]}\rightarrow Y^{[2]}$. Next let us
introduce the following $H$-valued one form $\ax$
\eq \label{cancon}
\ax =\sum_{\alpha} \rho_{\alpha}
{r_{\alpha}^{[2]}}^*\vvphi_{12}^{-1}\left(\overline{f}\,d\,\overline{f}^{-1}\right)\,.
\en
We easily find that $$
l^{h*}\ax=Ad_h \ax + p_1^*\left(h \sum_{\alpha}\rho_{\alpha}
{r_{\alpha}^{[1]}}^*\vvphi^{-1}(d\,\vvphi(h^{-1})\right)\,.
$$
The Lie$(Aut(H))$-valued 1-form
$\sum_{\alpha}\rho_{\alpha}
{r_{\alpha}^{[1]}}^*\vvphi^{-1}\,d\,\vvphi$ is, due to (\ref{equivariance}),
well defined on $Y$. We set 
\eq \label{canA}
A= \sum_{\alpha}\rho_{\alpha}
{r_{\alpha}^{[1]}}^*\vvphi^{-1}\,d\,\vvphi - d
\en 
for the sought Lie$(Aut(H))$-valued 1-form
on $Y$. Using the cocycle property
of $\bar{f}$ and $\vvphi$ we easily have
\begin{proposition}\label{localconstr}
Formulas (\ref{cancon}) and (\ref{canA}) give a bundle gerbe connection.
\end{proposition}
Using (\ref{lastlast}) we obtain that the 2-curvature $(\kx, K)$ of the bundle gerbe
2-connection $(\ax,A)$ satisfies
\eq
\kx_{12}\+\kx_{23}\+\kx_{13}^r= T_{K_1}(\overline{f}^{-1})~.
\en
 
\sk
\subsection*{Connection on a lifting bundle gerbe}

\noindent \\
Let us now consider the example of a lifting bundle gerbe associated with
an exact sequence of groups 
(\ref{ext}) and  a $G$-principal bundle $P\rightarrow M$ on $M$.
In this case, for any given  connection $\bar A$ on $P$ we can construct a
connection on the lifting bundle gerbe.  
Let us choose a section $s\,$: 
Lie$(G) \rightarrow $Lie$(E)$; i.e a linear map such that $\pi\circ
s=id$. 
We first define $A=s(\bar A)$ and then consider the Lie$(E)$ valued one-forms on  
$P^{[2]}$ given by $A_1= p_1^* s(\bar A)$ 
and $A_2= p_2^* s(\bar A)$, were
$p_1$ and $p_2$ are respectively the projections onto the first and second factor 
of $P^{[2]}$.
We next consider the one-form $\ax$ on $\EE$ that on
$(p_1,p_2;e)\in\EE$ is given by  
\eq
\ax\equiv e{\mathcal A}_2e^{-1}+ede^{-1}-{\mathcal A}_1~,\label{adefm}
\en
here ${\mathcal A}_{1}=p^*(A_{1})$ and 
${\mathcal A}_{2}=p^*(A_{2})$,  with $p\,:\,\EE\rightarrow P^{[2]}$. 
It is easy to see that $\pi^*\ax=0$ and that therefore $\ax$ is Lie$(H)$
valued; moreover $(\ax, ad_A)$ is a 2-connection on $\EE$. 
Recalling that on $\EE$ we have $\vvphi_{(p_1, p_2; e)} = Ad_e$, 
it is now a straightforward check left to the reader to show that 
$(\ax, ad_A)$ is a connection on the lifting bundle gerbe.

\subsection*{Connection on a module}

\noindent\\
Let us start discussing the case of the canonical module $can=Aut(H)\times Y$
(see Section $6$). Let $(\ax, A)$ be a connection on our bundle gerbe
$(\EE,Y,M,\ff)$. The Lie$(Aut(H))$-valued one-form $A$ on $Y$ lifts
canonically to the connection $\tilde{\mathcal A}$ on $can$ defined, forall 
$(\eta,y)\in can$, by
$\tilde{\mathcal A}= \eta {\mathcal A} \eta^{-1} + \eta d
\eta^{-1}$. 
Let us consider the following diagram 
\eq
\xymatrix{
        & can_{1} \oplus \EE  \ar[d]_{\pi_{1}}
        \ar[dr]^{\pi_{\oplus}}\ar[r]^{\hspace {0.5cm}\pi_{2}}  & \EE &\\
        & can_{1} & can_1\EE \ar[r]^{\hspace {0.3cm}z} & can_{2}\,.
        }
\en
As in the case of the bundle gerbe connection we can consider whether the
Lie$(Aut(H))$-valued one-form $\tilde{\mathcal A}_1 + \vvxi(\ax)$ that
lives on $can_{1} \oplus \EE$ is the pull-back
under $\pi_{\oplus}$ of a one-form connection on $can_1\EE$. If this is the
case then we say that $\tilde{\mathcal A}_1$ and
$\ax$ are summable and we denote by $\tilde{\mathcal A}_1 \+
ad_{\axs}$ the resulting connection  on $can_1\EE$.
Let us recall that on $can$ we have
$\vvxi_{(\eta,y)} = Ad\circ\vvpsi_{(\eta, y)}$ 
with $\vvpsi_{(\eta, y)}(h)=\eta(h)$. It is
now easy to check that
$\tilde{\mathcal A}_1$ and $\ax$ are summable and that their sum equals
the pull-back under $z$ of the connection $\tilde{\mathcal A}_2$;
in formulae
\eq
\tilde{\mathcal A}_1 \+ ad_{\axs}=z^*\tilde{\mathcal A}_2\,.\label{ultim}
\en
We also have that equality (\ref{ultim}) is equivalent to the
summability condition (\ref{sum-con2}) for the bundle gerbe connection $\ax$.
Thus (\ref{ultim}) is a new interpretation of the summability condition
(\ref{sum-con2}).
\sk
We now discuss connections on an arbitrary  module
$(\QQ, \ZZ, \zz)$ associated with a bundle gerbe $(\EE, Y, M, \ff)$ with
connection $(\ax, A)$. There are two natural requirements that a left
connection ${\mathcal A}^D$ on the left $D$-bundle $\QQ$ has to
satisfy in order to be a module connection. The first one is that 
the induced connection $\widehat{{\mathcal A}^D}$ on $Aut(H)\times_D\QQ$ has to be equal
(under the isomorphism $\sigma$) to the connection $\tilde {\mathcal A}$ of
$can$. This condition reads
\eq
{{\mathcal A}^D}=\vvpsi{{\mathcal A}}\vvpsi^{-1}+\vvpsi d\vvpsi^{-1}\label{1cond}~,
\en
where in the l.h.s.  ${{\mathcal A}^D}$ is tought to be Lie$(Aut(H))$
valued.  In other words on $Y$ we require $\sigma^*\widehat{{\mathcal A}^D}=A$, where
$\sigma$ is the global section of $Aut(H)\times_D\QQ$. 

Next consider the diagram 
\eq
\xymatrix{
        & \QQ_{1} \oplus \EE  \ar[d]_{\pi_{1}}
        \ar[dr]^{\pi_{\oplus}}\ar[r]^{\hspace {0.5cm}\pi_{2}}  & \EE &\\
        & \QQ_{1} & \QQ_1\EE \ar[r]^{\hspace {0.3cm}z} & \QQ_{2}\,~.
        }
\en
We denote by ${\mathcal A}_1^D \+ \alpha(\ax)$  the well defined $D$-connection 
on $\QQ_1\EE$ that pulled back on $\QQ_1\oplus \EE$
equals $\pi_1^*{\mathcal A}_1^D +  \vvxi(\pi_2^*\ax)$. It is not difficult to see that
${\mathcal A}_1^D$ is indeed summable to $\ax$ if 
forall $h\in H$, $\al(T_{{\mathcal A}^D}(h))=
\al(T_{\vvps{\mathcal A}\vvps^{-1}+\vvps d\vvps^{-1}}(h))$.
This  summability condition is thus implied by (\ref{1cond}). 
The second requirement that 
${{\mathcal A}^D}$ has to satify in order to be a module connection is
\eq 
{\mathcal A}_1^D\+ \alpha(\ax)=z^* {\mathcal A}^D_2 \,.\label{mcon}
\en
These conditions imply the summability condition (\ref{sum-con2}) for the bundle 
gerbe connection $\ax$.

\sk
Concerning the $D$-valued curvature ${\mathcal K}^D = d{\mathcal A}^D + {\mathcal A}^D \wedge 
{\mathcal A}^D$ we have
\eq
{\mathcal K}_1^D \+ \alpha(\kx_{\axs})=z_{12}^*{\mathcal K}^D_2 ~.
\en
\sk
In terms of local data a gerbe connection consists of a collection of local $2$-connections 
$(\ax_{\alpha \beta}, A_{\alpha})$ on the local bibundles $\EE_{\alpha
\beta}\rightarrow O_{\alpha \beta}$. For simplicity we assume the covering
$\{O_{\alpha}\}$ to be a good one. The explicit relations
that the local maps $f_{\alpha \beta \gamma} :O_{\alpha \beta
\gamma}\rightarrow H$, $\vphi_{\alpha \beta}: O_{\alpha \beta}\rightarrow Aut(H)$
and the local representatives $A_{\alpha}$, $K_{\alpha}$, $a_{\alpha
\beta}$ and $k_{\alpha \beta}$ (forms on $O_\alpha$, 
$O_{\alpha \beta}$, etc.) satisfy are
\eq\label{localf}
f_{\alpha \beta \gamma} f_{\alpha \gamma \delta}=\vphi_{\alpha
\beta}(f_{\beta \gamma \delta})f_{\alpha \beta \delta}\,,
\en 
\eq\label{localphi}
\vphi_{\alpha \beta}\vphi_{\beta \gamma}= Ad_{f_{\alpha \beta
\gamma}}\vphi_{\alpha \gamma}\,,
\en
\eq\label{locala}
a_{\alpha \beta}+\vphi_{\alpha \beta}(
a_{\beta \gamma})= 
f_{\alpha \beta \gamma}a_{\alpha \gamma}f^{-1}_{\alpha \beta \gamma} +
f_{\alpha \beta \gamma}\,d\,f^{-1}_{\alpha \beta \gamma} +
T_{A_{\alpha}}(f_{\alpha \beta \gamma})\,,
\en
\eq\label{localA}
A_{\alpha}+ad_{a_{\alpha \beta}}=\vphi_{\alpha \beta} A_{\beta}\vphi^{-1}_{\alpha \beta} +
\vphi_{\alpha \beta}\,d\,\vphi^{-1}_{\alpha \beta}\,, 
\en
\eq\label{localk}
k_{\alpha \beta}+\vphi_{\alpha \beta}(
k_{\beta \gamma})= 
f_{\alpha \beta \gamma}k_{\alpha \gamma}f^{-1}_{\alpha \beta \gamma} +
T_{K_{\alpha}}(f_{\alpha \beta \gamma})
\en
and
\eq\label{localK}
K_{\alpha}+ad_{k_{\alpha \beta}}=\vphi_{\alpha \beta} K_{\beta}\vphi^{-1}_{\alpha \beta}\,. 
\en
\sk
\sk
\sk
\section{\bf Curving}

In this section we introduce the curving two form $\bb$. This is achieved
considering a gerbe stably equivalent to  $(\EE, Y, M, \ff)$. The resulting
equivariant $H$-valued 3-form $\hh$ is then shown to be given in terms of a form on $Y$.
This description applies equally well to the abelian case; there one can however impose 
an extra condition [namely the vanishing of (\ref{bkb})]. We also give an explicit 
general construction of the curving $\bb$ in terms of a partition of unity. This construction
depends only on the partition of unity, and in the abelian case it naturally reduces to the 
usual one that automatically encodes the vanishing of (\ref{bkb}).

Consider a bundle gerbe $(\EE, Y, M, \ff)$ with connection 
$(\ax,A)$ and curvature $(\kx_{\axs}, K_A)$ and an $H$-bibundle
$\NN \rightarrow Y$ with a $2$-connection $(\cc, A)$. Then we have a stably equivalent gerbe 
$(\NN_1^{-1}\EE\NN_2 , Y, M, \vvphi_1^{-1}(\ff))$ with connection $(\thetaa,
A^{r_1})$ given by
\eq
\thetaa = \cc_1^{r_1} \+ \ax \+ \cc_2\,.
\en 
Also we can consider a $K_A$-equivariant horizontal $2$-form $\bb$ on
$\NN$. Again on the bibundle $\NN_1^{-1}\EE\NN_2 \rightarrow Y^{[2]}$ we get a 
well defined 
$K^{r_1}_A$-equivariant horizontal $2$-form
\eq
\tdeltt = \bb_1^{r_1} \+ \kx_{\axs} \+ \bb_2\,.\label{bkb}
\en 
Contrary to the abelian case we cannot achieve $
\tdeltt=0$, unless $K_A$
is inner (remember $\tdeltt$ is always $K_A^{r_1}$-equivariant). 
Next we consider the equivariant horizontal $H$-valued $3$-form $\hh$
on $\NN$  given by 
\eq
\hh = D_{\ccs} \bb\,.\label{definitH}
\en
Because of the Bianchi identity $dK_A + [A,K_A] =0$ this is indeed an equivariant form on 
$\NN$. Obviously the horizontal form $\vvphi^{-1}(\hh)$ is invariant
under the left $H$-action
\eq
{l^h}^*\vvphi^{-1}(\hh) = \vvphi^{-1}(\hh)
\en and therefore it projects to a well defined form on $Y$.

Using now the property of the covariant derivative (\ref{sumforms}) and the 
Bianchi identity (\ref{Bianchi}) we can 
write
\eq 
\hh_1^r \+ \hh_2 = D_{\thetaas} \tdeltt \,.
\en

Finally from (\ref{Dsquare}) we get the Bianchi identity for $\hh$
\eq
D_{\ccs}\hh = [\kx_{\ccs}, \bb] + T_{K_A}(\kx_{\ccs}) - T_{K_A}(\bb)\,.
\en

{}For the rest of this section we consider the special case where 
$\NN$ is a trivial bibundle  with  global central
section $\bar \sigma$ and with $2$-connection given by $(\cc, A)$, 
where $\cc$ is canonically given by $\bar \sigma$, 
$$\cc = \bar{\sigma} d \bar{\sigma}^{-1} + T_A(\bar{\sigma})~.$$
Since the only $H$-bibundle $\NN\rightarrow Y$ that we can canonically
associate to a generic bundle gerbe is the trivial one 
(see Proposition $\ref{D-H-triv}$), the special case where $\NN$ is trivial
seems quite a  natural case.

In terms of local data curving is a collection 
$\{\bb_{\alpha}\}$ of ${K_{\alpha}}$-equivariant horizontal two forms on 
trivial $H$-bibundles
$O_{\alpha}\times H \rightarrow O_{\alpha}$. Again we assume the covering
$O_{\alpha}$ to be a good one and write out explicitly the relations
to which the local representatives of $b_{\alpha}$ and $h_{\alpha}$
(forms on $O_\alpha$) are subject:
\begin{eqnarray}
&&k_{\alpha \beta}+ \vphi_{\alpha \beta}(b_\beta)= 
b_{\alpha} + \delta_{\alpha \beta} \label{localb}\,,\\
&&\delta_{\alpha \beta}+\vphi_{\alpha \beta}(
\delta_{\beta \gamma})=
f_{\alpha \beta \gamma}\delta_{\alpha \gamma}f^{-1}_{\alpha \beta \gamma}  +
T_{\nu_{\alpha}}(f_{\alpha \beta \gamma})\label{localdelta}\,,\\
&&\nu_{\alpha} \equiv K_{\alpha} - ad_{b_{\alpha}}\,,\\
&&h_{\alpha} = d\, b_{\alpha}- T_{A_{\alpha}}(b_{\alpha})\label{hlocal}\,,\\
&&\vphi_{\alpha \beta}(h_\beta)= h_{\alpha}+ d \delta_{\alpha \beta}+ [a_{\alpha
\beta},\delta_{\alpha \beta}] + T_{K_{\alpha}}(a_{\alpha \beta}) - 
T_{A_{\alpha}}(\delta_{\alpha \beta})\label{localh} 
\end{eqnarray}
and the Bianchi identity
\eq\label{localBianchi}
d\, h_{\alpha} + T_{K_A}(b_{\alpha}) = 0\,.\en
Here we introduced $\delta_{\alpha \beta} = \vphi_{\alpha}(\tilde
\delta_{\alpha \beta})$\,.
Equations 
(\ref{localf})-(\ref{localK}) and (\ref{localb})-(\ref{localBianchi})
are the same as those listed after Theorem $10.1$ in \cite{Breen-Messing}.

We now consider the case $Y= \sqcup O_\al$; this up to stable
equivalence is always doable. Given 
a partition of unity $\{\rho_{\alpha}\}$
subordinate to the covering $\{O_{\alpha}\}$ of $M$,  we have a
natural choice 
for the $H$-valued curving
$2$-form
$\bb$ on $\sqcup O_\al\times H$ . It is the pull-back under the projection
$\sqcup O_\al\times H
\rightarrow \sqcup O_\al$ of the $2$-form 
\eq \label{partb}
\sqcup \sum_{\beta} \rho_{\beta} k_{\alpha \beta}
\en
on $Y=\sqcup O_\al$.
In this case we have for the local $H$-valued $2$-forms $\delta_{\alpha \beta}$ 
the following expression
\begin{eqnarray}
\delta_{\alpha \beta}&=& \sum_{\gamma}\rho_{\gamma}(
f_{\alpha \beta \gamma}k_{\alpha \gamma}f^{-1}_{\alpha \beta \gamma} -
k_{\alpha \gamma} +
T_{K_{\alpha}}(f_{\alpha \beta \gamma}))\nonumber\\
&=& \sum_{\gamma} \rho_{\gamma}(k_{\alpha \beta} + \vphi_{\alpha \beta}
(k_{\beta \gamma})-k_{\alpha \gamma})\,.
\end{eqnarray}
We can now use Proposition \ref{localconstr} together with
(\ref{partb}) in order to
explicitly construct from the \v Cech cocycle $(\ff, \vvphi)$ 
an $H$-valued $3$-form $\hh$.

We conclude this final section by grouping together the global
cocycle formulae that imply all the local expressions (\ref{localf})-(\ref{localK}) 
and (\ref{localb})-(\ref{localBianchi}),
\vspace{.1 cm}
$$
\ff_{124}{}{\,} \mbox{\boldmath $\vphi$}_{12}(\ff_{234})
=\ff_{134}{}{\,}\ff_{123}\,,
$$
\vspace{-1.2 cm}
\begin{flushright}
(\ref{cocycleglobal})
\end{flushright}
\vspace{.1 cm}
$$
\ax_{12} \+ \ax_{23}=f^*\ax_{13}\,,
$$
\vspace{-1.2 cm}
\begin{flushright}
(\ref{sum-cond1})
\end{flushright}
\vspace{.1 cm}
$$
\tdeltt = \bb_1^{r_1} \+ \kx_{\axs} \+ \bb_2\,,
$$ 
\vspace{-1.2 cm}
\begin{flushright}
(\ref{bkb})
\end{flushright}
\vspace{.1 cm}
$$
\hh = D_{\ccs} \bb~.
$$
\vspace{-1.2 cm}
\begin{flushright}
(\ref{definitH})
\end{flushright}

\sk\sk\sk
\section*{\bf Acknowledgements}
We have benefited from discussions with L. Breen, D. Husemoller, 
A. Alekseev, L. Castellani,
J. Kalkkinen, J. Mickelsson, R. Minasian, D. Stevenson and R. Stora.
\sk
\sk\sk

\sk

\end{document}